\documentclass{article}

\usepackage[numbers]{natbib}         
\usepackage[colorlinks]{hyperref}    
\usepackage[english]{babel} 
\usepackage{amssymb}
\usepackage{amsmath}
\usepackage{txfonts}
\usepackage{mathdots}
\usepackage[classicReIm]{kpfonts}
\usepackage[dvips]{graphicx} 
\usepackage[a4paper, portrait, margin=1in]{geometry}
\usepackage{tabularx}
\usepackage{longtable}
\usepackage{multirow}
\usepackage{booktabs}
\usepackage[labelsep=period]{caption}
\usepackage{makecell}
\usepackage{lscape}
\begin{document}
	\setlength{\parindent}{0pt}
	\setlength{\parskip}{1ex}
	
	\textbf{\Large A Review of Artificial Intelligence in Brachytherapy}
	
	\bigbreak

	Jingchu Chen$^{1,2}$,Richard L.J. Qiu$^{1}$, Tonghe Wang$^{3}$, Shadab Momin$^{1}$ and Xiaofeng Yang$^{1*}$

	1Department of Radiation Oncology and Winship Cancer Institute, Emory University, Atlanta, GA 30308
	
	2School of Mechanical Engineering, Georgia Institute of Technology, GA, Atlanta, USA
	
	3Department of Medical Physics, Memorial Sloan Kettering Cancer Center, New York, NY 10065

	\bigbreak
	\bigbreak
	\bigbreak

	\textbf{*Corresponding author: }
	
	Xiaofeng Yang, PhD
	
	Department of Radiation Oncology
	
	Emory University School of Medicine
	
	1365 Clifton Road NE
	
	Atlanta, GA 30322
	
	E-mail: xiaofeng.yang@emory.edu

	\bigbreak
	\bigbreak
	\bigbreak
	\bigbreak
	\bigbreak
	\bigbreak

	\textbf{Abstract}

	Artificial intelligence (AI) has the potential to revolutionize brachytherapy's clinical workflow. This review comprehensively examines the application of AI, focusing on machine learning and deep learning, in facilitating various aspects of brachytherapy. We analyze AI's role in making brachytherapy treatments more personalized, efficient, and effective. The applications are systematically categorized into seven categories: imaging, preplanning, treatment planning, applicator reconstruction, quality assurance, outcome prediction, and real-time monitoring. Each major category is further subdivided based on cancer type or specific tasks, with detailed summaries of models, data sizes, and results presented in corresponding tables. This review offers insights into the current advancements, challenges, and the impact of AI on treatment paradigms, encouraging further research to expand its clinical utility.
	
	
	\bigbreak
	\bigbreak

	\noindent 
	\section{ INTRODUCTION}
	
	Brachytherapy is a form of internal radiation therapy that is delivered with either low dose rate (LDR) or high dose rate (HDR). It involves the direct placement of radioactive sources within or near the tumor via applicators. Brachytherapy plays a crucial role in radiation therapy due to its ability to deliver high and conformal radiation doses to the tumor with reduced dose to adjacent organ-at-risk (OARs), thereby offering an advantageous therapeutic ratio. \cite{RN104, RN122, RN103, RN123, RN120}
	
	After patient consultation and consent for treatment, both LDR and HDR, forms of brachytherapy procedures, may contain several key steps: (a) Preplanning diagnostic imaging; (b) Surgical insertion of needles, applicators, or catheters; (c) Treatment planning imaging and verification; (d) Image registration and segmentation; (e) Applicators/catheters reconstruction; (d) Treatment planning; (f) Quality assurance (QA); (g) Treatment delivery; (h) Patient follow up. 
	
	Overall workflow for brachytherapy procedures can be labor and resource intensive for personnel involved from different disciplines. However, efficiency and efficacy of brachytherapy procedures is highly dependent on the clinician's skills and level of experience owing to their involvement at various points in the workflow including diagnostics, implantation, and treatment planning. With the goal of further improving the efficiency and addressing the challenges present in brachytherapy procedures, studies have integrated artificial intelligence (AI) to facilitate these procedures. In recent years, it has become clear that AI, which could be considered the fourth industrial revolution, is emerging as a transformative force across various sectors, including healthcare. \cite{RN126, RN124} Radiation oncology and medical physics, disciplines at the forefront of integrating cutting-edge scientific and technological innovations, are increasingly exploring the potential of AI to revolutionize treatment paradigms. \cite{RN107, RN108, RN125} Within the realm of AI, Machine learning (ML) relies on statistical models to learn from previous data and make predictive decisions, which can be useful in selecting brachytherapy applicators and predicting outcomes. \cite{RN82, RN127, RN56, RN83, RN128} Deep learning (DL), a subset of ML, uses neural networks such as Convolutional Neural Network (CNN) \cite{RN91} and Generative Adversarial Network (GAN) \cite{RN105} to execute complex image analysis tasks \cite{RN106}, which are fundamentally involved in most of the brachytherapy workflow. \cite{RN129}
	
	Figure 1 shows the number of studies that utilized AI for different brachytherapy purposes from 2015 up to August 2024. After a significant increase in the number of studies on AI applications in brachytherapy up to 2020, there is a stagnation in the growth of these studies in the subsequent years (Figure 1). Given the rapid advancements and the potentials of AI to refine and redefine brachytherapy treatment workflow, a comprehensive review of the current and emerging applications of AI in the context of brachytherapy is both timely and pivotal to encourage more researchers to study brachytherapy and to provide physicians with an overview of the current state of AI in brachytherapy. Therefore, we collected recent developments in the applications of AI in different brachytherapy procedures to provide a detailed analysis of the potentials of AI in leading to more personalized, efficient, and effective brachytherapy treatments.
	
	\begin{figure}
		\centering
		\noindent \includegraphics*[width=6.50in, height=4.20in, keepaspectratio=true]{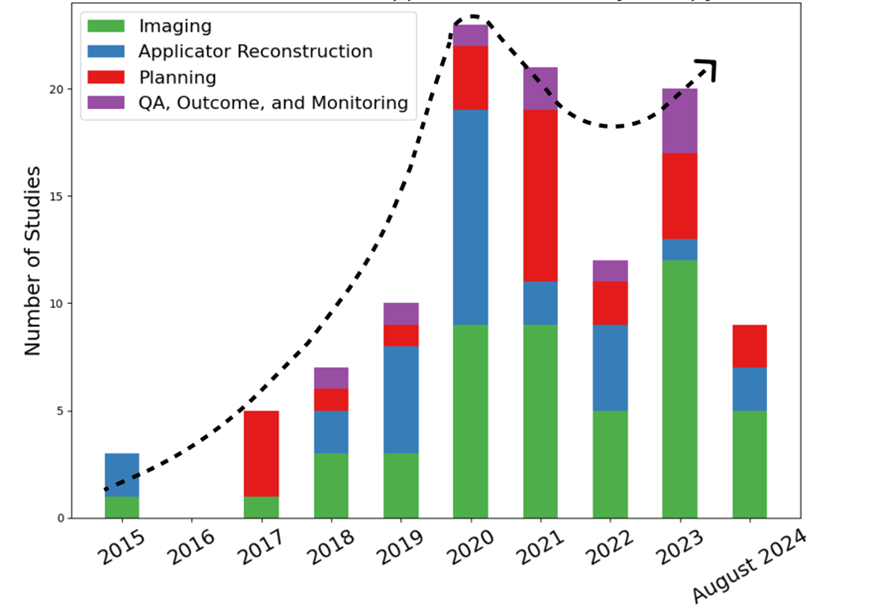}
		
		\noindent Figure 1. Overview of the number of studies in applications of AI in brachytherapy from 2015 to August 2024 with an approximate dotted trendline. 'Imaging' combines the image registration, image segmentation, and other imaging-related tasks. 'Planning' includes both preplanning and treatment planning. 'Applicator reconstruction' includes both prostate and gynecologic cancer-related applicator reconstructions including needles, catheters, and seeds. 'QA, Outcome, and Monitoring' combined the remaining three categories because of their relatively small number of studies.
	\end{figure}

	The literature search was conducted in August 2024, initially retrieving 205 papers from PubMed using the keywords "brachytherapy", combined with "artificial intelligence", "deep learning", or "machine learning". Additional papers were supplemented from The American Association of Physicists in Medicine (AAPM) and Google Scholar by searching the same keywords. A total of 111 studies were selected and thoroughly reviewed. The innovative approaches used to develop each unique AI model, along with their corresponding performances, are presented and categorized by different clinical procedures in brachytherapy workflow, with further sub-categorization based on specific organs or detailed applications.
	
	The studies are categorized into 7 sections: imaging (section 2), preplanning (section 3), treatment planning (section 4), applicator reconstruction (section 5), QA (section 6), outcome prediction (section 7), and real-time monitoring (section 8). The detailed percentage distribution of these studies is illustrated in Figure 2. 
	
	\begin{figure}
		\centering
		\noindent \includegraphics*[width=6.50in, height=4.20in, keepaspectratio=true]{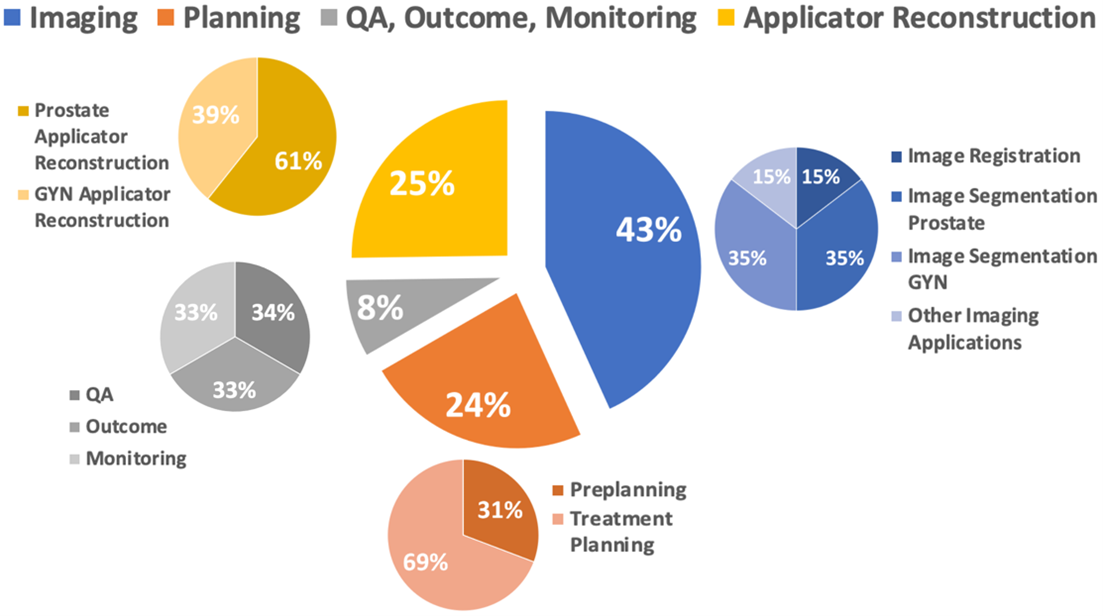}
		
		\noindent Figure 2. Percentage pie chart of applications of AI in different procedures in brachytherapy.
	\end{figure}

	Below is a summary of several common evaluation methods used in the literature for evaluating the performance of the AI models. 
	
	1.	The Dice Similarity Coefficient (DSC) is the most used metric for measuring the overlap ratio of between the automatic (A) and ground truth (B) contours. 
	\begin{equation} 
		DSC=\frac{2\times (A\cap B)}{|A|+|B|},
	\end{equation} 

	2.	Hausdorff Distance (HD) measures the maximum distance between points in the predicted segmentation and points in the ground truth images, serving as a measure of dissimilarity. HD95, commonly used in many studies, disregards outliers by considering only the 95th percentile of HD. 
	\begin{equation} 
		HD95=\max_{k95\%}\{\sup_{a\in A}\inf_{b\in B}d(a,B),\sup_{b\in B}\inf_{a\in A}d(A,b)\},
	\end{equation}
	where: a are the points belong to image set A, and b are the points belong to image set B.
	
	3.	Mean Surface Distance (MSD) compares the average difference between the surface of automatic segmentation (A) and ground truth contours (B).
	\begin{equation} 
		MSD=\frac{1}{A}\Sigma_{a\in A}\min_{b\in B}|a-b|+\frac{1}{B}\Sigma_{b\in B}\min_{a\in A}|b-a|,
	\end{equation}
	where: a are the points belong to image set A, and b are the points belong to image set B.
	
	\noindent 
	\section{Imaging}

	Many modern imaging models use CNN \cite{RN91} or fully convolutional networks (FCN) \cite{RN92}, which incorporate spatial information to perform pixel-wise classification for computer vision tasks. Ronneberger et al. \cite{RN90} built the first U-Net structure based on FCN specifically for biomedical image segmentation. The U-Net model specializes in biomedical image segmentation because of its incorporation of data augmentation with elastic deformation, which reduces the need for large training dataset. This model also addresses tissue deformation variability through elastic deformation data augmentation. Several models use U-Net as a backbone, enhancing segmentation results with additional multiple skip connections between encoder levels and self-adapting frameworks, such as Unet++ \cite{RN93} and nnU-Net (no-new-net) \cite{RN94}. Additionally, U-Net can be combined with Transformers, such as the TransUNet \cite{RN96} and UNETR (UNEt TRansformers) \cite{RN95}, which use local semantic and texture information while incorporating long-range dependencies among pixels \cite{RN97}. Since imaging is a crucial element in brachytherapy, the neural network-based models and their variations are extensively employed to perform image registration, image segmentation, and other applications. 
	
	\noindent 
	\subsection{Image Registration}
	
	Image registration is the process of aligning multimodality medical images or the single modality medical images between different treatment fractions in brachytherapy. Table 1 provides an overview of the methods and results from current AI-based image registration implementations. 
	
	Gynecologic (GYN) brachytherapy typically involves multiple treatment fractions, during which organ deformation occurs due to varying bladder and rectum filling, applicator insertion, and inter-fractional tumor changes \cite{RN98}. Figure 3 illustrates the inter-fractional changes of the gross tumor volume (GTV), which is labeled in red, in GYN brachytherapy. Organ deformations cause dosimetric uncertainty for the target and OARs, making image registration necessary. The transformation in image registration can be subdivided into rigid, affine, and deformable, while the registration method can be either intensity-based or geometric-based.

	\begin{figure}
		\centering
		\noindent \includegraphics*[width=6.50in, height=4.20in, keepaspectratio=true]{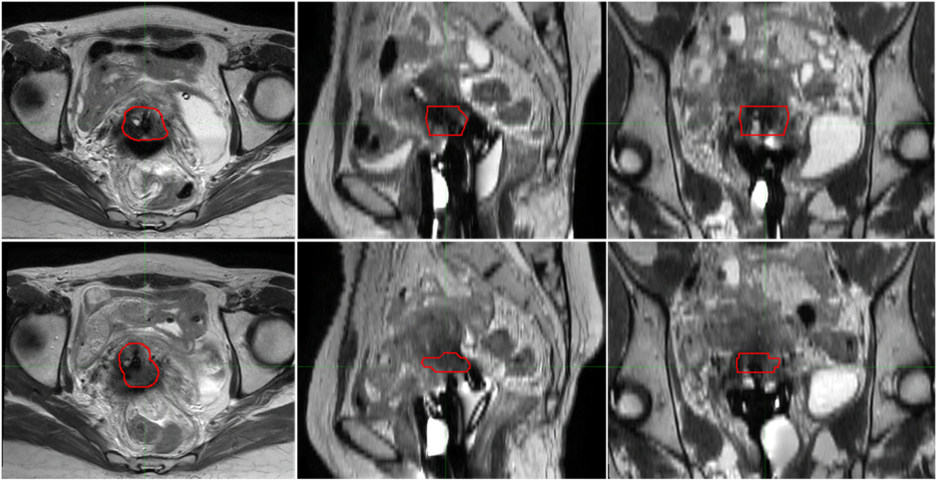}
		
		\noindent Figure 3: Inter-fractional changes in gynecologic brachytherapy. The red circle indicates the GTV. The images display axial (left column), sagittal (middle column), and coronal (right column) views, demonstrating the variations observed across two treatment fractions with one week interval (upper and lower rows).
	\end{figure}

	To address inter-fractional changes of organs involved in GYN brachytherapy, Salehi et al. \cite{RN17} developed a deep learning-based non-rigid deformable image registration algorithm (DIRNet) for aligning CT images of the cervix and OARs. By fixing the bony structures and applying deformed binary masks for the organs, the study showed that DIRNet achieved comparable results in the DSC and significantly better Jaccard distance (JI) – the ratio of the intersected region relative to the union of the automatic and ground truth contours - and MSD than the conventional intensity-based image registration (SimpleElastix), as detailed in Table 1. Besides deformable image registration, rigid registration can be performed based on applicator geometries. Ecker et al. \cite{RN18} combined deep learning-based applicator segmentation with existing rigid registration methods to automate the registration between MR-guided GYN brachytherapy fractions. A 3D UNETR model was used to segment applicators with a DSC of 0.70 ± 0.07 and served as the reference for rigid registration. The mean distance error (MDE) of registration using the predicted segmentation was 2.7 ± 1.4 mm, which is relatively high compared to the error using ground truth segmentation, which was only 0.7 ± 0.5 mm. Although the current registration error remains above the desired registration error of 2 mm, this discrepancy is likely due to the limited segmentation accuracy of applicators. The study demonstrated that if a segmentation method that produces similar results as the ground truth is used, the registration error can be reduced, achieving an automated registration process significantly faster than manual methods. 
	
	In prostate brachytherapy, image registration is crucial for fusing organ information between multiple image modalities. Transrectal ultrasound (TRUS) is often used for guiding the insertion of needles or catheters in prostate brachytherapy, while MRI offers greater soft tissue contrasts compared to TRUS. The incorporation of knowledge from MRI into TRUS images can thus provide additional soft tissue guidance. However, registering MRI to TRUS images remains challenging due to the limited availability of ground truth deformation of the prostate. 
	
	To address this challenge, several different AI-based methods are provided. Zeng et al. \cite{RN19} developed a fully automated deep learning system based on a weakly supervised method, which trains models using only partially labeled data. The entire workflow could be summarized into three steps: initial segmentation on TRUS and MRI using two FCNs, affine registration using a 2D CNN, and non-rigid registration using 3D U-Net based network. The initial segmentation results produced reliable results for the registration, where the DSC = 0.88 ± 0.05 and 0.92 ± 0.03 for the MRI and for TRUS. The affine and non-rigid registration methods were followed using MRI-TRUS labels (SR-L) and MRI-TRUS images (SR-I) as inputs. Overall, the model produced the best DSC, mean target registration error, mean MSD, and HD using the SR-L input method, with DSC = 0.91 ± 0.02, target registration error = 2.53 ± 1.39 mm, MSD = 0.88 mm, and HD = 4.41 mm. Using the labels as input and deep-learning methods for initialization, Zeng’s approach showed high accuracy in automatic registration of MRI-TRUS images of prostate. 
	
	Besides the label-based method, Chen et al. \cite{RN20} provided a segmentation-based method, which used 3D V-Net models to segment the prostate after catheter insertion on MR and US images, align the centroids, and probability maps to predict deformable displacement fields. Despite the presence of catheter artifacts in the images, the model achieved a DSC of 0.87 ± 0.05, a Center of Mass distance error of 1.7 ± 0.89 mm, an HD of 7.21 ± 2.07 mm, and an MSD of 1.61 ± 0.64 mm. The segmentation-based method provided slightly inferior results than the label-based method, likely due to the results of initial segmentation (DSC values of 0.86 ± 0.05 and 0.90 ± 0.03 on MR and US) and utilization of hierarchical information of anatomy rather than image intensities. While current studies have shown positive outcomes when integrating MRI into TRUS-based workflows, using MRI-registered contours, whether rigid, semi-rigid, or deformable, may still lead to significant dose under-coverage. \cite{RN132}
	
	\begin{table}[]
		\caption{AI applications in image registration}
		\label{tab:my-table}
		\resizebox{\columnwidth}{!}{%
			\begin{tabular}{|p{3cm}|p{2cm}|p{2.5cm}|p{2cm}|p{2cm}|p{7cm}|p{2cm}|}
				\hline
				Cancer   Site & Image   Modality & Registration   Method & Number   of Patients & Model & Result   Summary & Citation \\ \hline
				GYN & CT & Non-rigid & 57 & DIRNet & \begin{tabular}{p{6.5cm}}Mean MSD (mm): Model = 1.61 ± 0.46, 1.17 ± 0.15, 1.06 ± 0.42; SimpleElastix = 2.94 ± 0.78, 3.26 ± 0.74, 3.04 ± 1.50 (cervix, bladder, rectum).\\ Mean MSD (mm): Model = 1.61 ± 0.46, 1.17 ± 0.15, 1.06 ± 0.42; SimpleElastix = 2.94 ± 0.78, 3.26 ± 0.74, 3.04 ± 1.50 (cervix, bladder, rectum).\\ Mean JI: Model = 86 ± 4\%, 93 ± 1\%, 88 ± 4\%; SimpleElastix = 71 ± 8\%, 83 ± 4\%, 67 ± 11\% (cervix, bladder, rectum). \end{tabular} & \cite{RN17} \\ \hline
				GYN & MRI & Rigid & 56 & 2D   U-Net and 3D UNETR & MDE   between dwell positions = 2.7±1.4 mm. & \cite{RN18} \\ \hline
				Prostate & MRI-TRUS & Affine   and non-rigid & 36 & FCN,   2D CNN, and 3D U-Net & DSC =   0.91 ± 0.02, target registration error = 2.53 ± 1.39 mm, MSD = 0.88 mm, and   HD = 4.41 mm. & \cite{RN19} \\ \hline
				Prostate & MRI-TRUS & Affine   and non-rigid & 32 & FCN   and RNN & DSC =   0.90 ± 0.04, target registration error = 2.77 ± 1.40 mm. & \cite{RN142} \\ \hline
				Prostate & MRI-TRUS & Rigid & 121 & 3D   V-Net and Probability Maps & DSC =   0.87 ± 0.05, Center of Mass distance error = 1.7 ± 0.89 mm, HD = 7.21 ± 2.07   mm, and MSD = 1.61± 0.64 mm. & \cite{RN20} \\ \hline
				Prostate & MRI-TRUS & Deformable & 642 & Weakly-supervised   Volumetric Registration & DSC =   0.873 ± 0.113, HD = 4.56 ± 1.95 mm, and MSD = 0.053 ± 0.026 mm. & \cite{RN130} \\ \hline
				Prostate & MRI-TRUS & Rigid & 662 & Attention-Reg & DSC = 0.82 ± 0.06   and Surface Registration Error = 5.99 ± 3.52 mm. & \cite{RN143} \\ \hline
			\end{tabular}%
		}
	Note: Abbreviations: RNN (recurrent neural network).
	\end{table}
	
	\noindent 
	\subsection{Image Segmentation}
	
	Image segmentation involves defining various target volumes that require treatment and OARs that require sparing during brachytherapy. It serves as the foundation for various tasks in the brachytherapy workflow including treatment planning. We summarized the AI segmentation methods and results for GYN-related tasks in Table 2 and for prostate-related tasks in Table 3.
	
	\noindent 
	\subsubsection{GYN}
	
	The GYN-related segmentation studies focused on segmenting the GTV, the high-risk clinical target volume (HR-CTV) which extends from the GTV to account for possible microscopic spread of cancer, and the OARs including bladder, rectum, sigmoid, and small intestine. As shown in Figure 4, various structures need to be considered for GYN brachytherapy patients. 
	
	\begin{figure}
		\centering
		\noindent \includegraphics*[width=6.50in, height=4.20in, keepaspectratio=true]{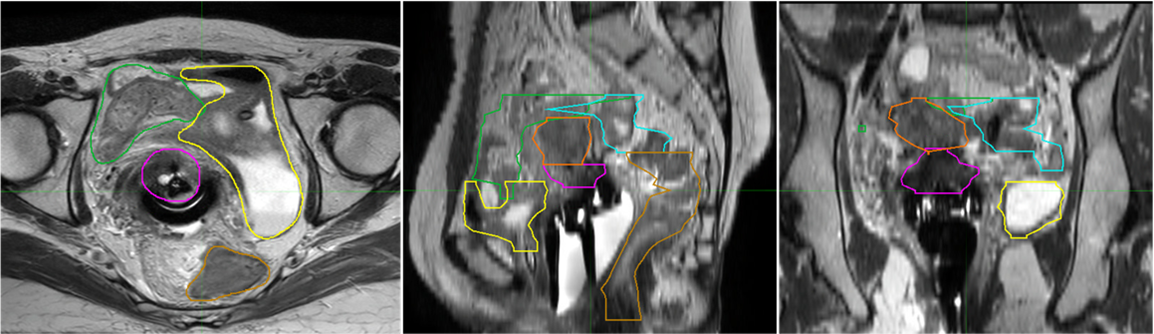}
		
		\noindent Figure 4: Image segmentation on MRI for a GYN brachytherapy patient, showing axial, sagittal, and coronal views. The segmented structures are indicated by different colors: magenta for the cervix, yellow for the bladder, green for the bowel, orange for the uterus, gold for the rectum, and blue for the recto sigmoid.
	\end{figure}
	
	Incorporating MRI into image-guided brachytherapy treatment planning enhances the clarity of targets and OARs. However, contouring organs on MRI is a labor-intensive process, especially problematic when performed while the patient is immobilized with the applicator inserted during brachytherapy. For this reason, several studies aim at developing an automated method for GYN organ segmentation on MRI post applicator insertion. Yoganathan et al. \cite{RN7} trained two deep CNN architectures, a residual neural network (ResNet50) and an inception residual network (InRN) to automatically segment the GTV, HR-CTV, and the OARs on MRI scans for patients with Foleys catheter inserted. The GTV often appears in non-uniform shapes and variable sizes, which makes it challenging to identify. The study implemented a 2.5D method that utilizes axial, sagittal, and coronal views to include additional contextual spatial information. Combining 2.5D model with InRN network architecture yielded optimal segmentation results as detailed in Table 2. However, MRI scans exhibit variations due to different clinical setting, such as different scanners vendors, scanning parameters, and applicator types. To enhance model robustness and adaptability to various applicators and MRI scanners, two studies incorporated diverse settings in their training data, aiming to develop a more generalized model. Zabihollahy et al. \cite{RN3} trained a 2-step CNN (3D Dense U-Net) with different MRI setups: MR1 with a repetition time (TR) of 2600 ms and time to echo (TE) of 95 ms, and MR2 with a TR of 3500 ms and TE of 97 ms. Ni et al. \cite{RN134} fine-tuned a pre-trained model with diverse training groups mixing different MRI scanners (1.5T Siemens Espree and 3T Siemens Verio) and brachytherapy applicators (tandem and ring, Syed-Neblett template, Venezia applicator). The optimal segmentation results of OARs from both studies suggest that training with mixed data improves generalizability, making the models more viable for future clinical implementation. The rapid segmentation of AI models can also facilitate MRI-guided brachytherapy planning. Kim et al. \cite{RN139} developed a dual convolution-transformer U-Net (DCT-UNet) that provided HR-CTV and OAR segmentation along with a real-time active needle tracking function to enhance the efficiency of MRI-guided brachytherapy procedures. The model is also incorporated into the treatment planning system (TPS), to assist radiation oncologists in precisely placing catheters under MRI guidance. The approach involved deformably registering the MRI from the primary treatment planning image with contours (T2SPACE) to the MRI acquired during the procedure (T2QM), enabling accurate and real-time tracking for improved guidance. The model achieved high DSC scores for OAR segmentations, despite having less optimal results for the HR-CTV on T2QM. This demonstrates its potential for improving catheter placement in MRI-guided brachytherapy.
	
	CT has been extensively used in image-guided adaptive brachytherapy, facing similar time-constraint challenges as MRI. Several studies have employed neural network-based models to simultaneously segment HR-CTV and OARs on CT images \cite{RN5, RN39, RN6, RN40, RN137, RN41, RN135}. Li et al. \cite{RN6}, Duprez et al. \cite{RN5}, and Xue et al. \cite{RN137} employed nnU-Net-based models to segment HR-CTV and OARs across various configurations, including 2D U-Net, 3D U-Net, and 3D Cascade U-Net. Unlike regular U-Net, the nnU-Net handles training parameters autonomously for new tasks and each segmentation task uses the best corresponding architecture. Li and Duprez’s studies produced similar results in terms of DSC, HD95, and MSD with the 3D Cascade U-Net configuration in nnU-Net. However, Xue’s study incorporated a prompt-based feature, which allows users to define a box around the target, outperforming the original nnU-Net model with a remarkably high DSC of 0.96 for HR-CTV and 0.91 for the rectum. All three studies found no statistically significant dosimetric differences between the manual and generated contours. These findings support the incorporation of a prompt-based feature in segmentation models to improve the results of complex structure segmentation, like the HR-CTV.
	
	\begin{table}[]
		\caption{AI-based GYN-related segmentation results}
		\label{tab:my-table}
		\resizebox{\textwidth}{!}{%
			\begin{tabular}{|l|l|l|l|llllll|l|}
				\hline
				\multirow{2}{*}{Image   Modality} & \multirow{2}{*}{Number   of Patients} & \multirow{2}{*}{Model} & \multirow{2}{*}{Evaluation   Parameters} & \multicolumn{6}{l|}{Results} & \multirow{2}{*}{Citation} \\ \cline{5-10}
				&  &  &  & \multicolumn{1}{l|}{HR-CTV} & \multicolumn{1}{l|}{Bladder} & \multicolumn{1}{l|}{Rectum} & \multicolumn{1}{l|}{Sigmoid   Colon} & \multicolumn{1}{l|}{Small   Intestine} & GTV &  \\ \hline
				\multirow{2}{*}{MRI} & \multirow{2}{*}{39} & \multirow{2}{*}{ResNet50   and InRN} & DSC & \multicolumn{1}{l|}{0.85 ±   0.06} & \multicolumn{1}{l|}{0.90 ±   0.05} & \multicolumn{1}{l|}{0.76 ±   0.07} & \multicolumn{1}{l|}{0.65 ±   0.12} & \multicolumn{1}{l|}{0.54 ±   0.12} & 0.62 ±   0.14 & \multirow{2}{*}{\cite{RN7}} \\ \cline{4-10}
				&  &  & HD95   (mm) & \multicolumn{1}{l|}{4.87 ±   2.19} & \multicolumn{1}{l|}{6.28 ±   3.42} & \multicolumn{1}{l|}{8.20 ±   4.07} & \multicolumn{1}{l|}{20.44   ± 11.70} & \multicolumn{1}{l|}{22.3 ±   13.66} & 6.83 ±   2.89 &  \\ \hline
				\multirow{4}{*}{MRI} & \multirow{4}{*}{129   for MR1 and 52 for MR2} & \multirow{4}{*}{3D   Dense U-Net} & DSC   (MR1) & \multicolumn{1}{l|}{-} & \multicolumn{1}{l|}{0.93 ±   0.04} & \multicolumn{1}{l|}{0.87 ±   0.03} & \multicolumn{1}{l|}{0.80 ±   0.10} & \multicolumn{1}{l|}{-} & - & \multirow{4}{*}{\cite{RN3}} \\ \cline{4-10}
				&  &  & DSC   (MR2) & \multicolumn{1}{l|}{-} & \multicolumn{1}{l|}{0.94 ±   0.05} & \multicolumn{1}{l|}{0.88 ±   0.04} & \multicolumn{1}{l|}{0.80 ±   0.05} & \multicolumn{1}{l|}{-} & - &  \\ \cline{4-10}
				&  &  & HD95   (mm) (MR1) & \multicolumn{1}{l|}{-} & \multicolumn{1}{l|}{4.18 ±   0.52} & \multicolumn{1}{l|}{2.54 ±   0.41} & \multicolumn{1}{l|}{5.03 ±   1.31} & \multicolumn{1}{l|}{-} & - &  \\ \cline{4-10}
				&  &  & HD95   (mm) (MR2) & \multicolumn{1}{l|}{-} & \multicolumn{1}{l|}{2.89 ±   0.33} & \multicolumn{1}{l|}{2.24 ±   0.40} & \multicolumn{1}{l|}{3.28 ±   1.08} & \multicolumn{1}{l|}{-} & - &  \\ \hline
				\multirow{6}{*}{MRI} & \multirow{6}{*}{136} & \multirow{6}{*}{nnU-Net   and transfer learning} & \begin{tabular}[c]{@{}l@{}}vDSC\\    \\ (US)\end{tabular} & \multicolumn{1}{l|}{-} & \multicolumn{1}{l|}{0.93 ±   0.04} & \multicolumn{1}{l|}{0.87 ±   0.06} & \multicolumn{1}{l|}{0.7 ±   0.2} & \multicolumn{1}{l|}{0.7 ±   0.1} & - & \multirow{6}{*}{\cite{RN134}} \\ \cline{4-10}
				&  &  & \begin{tabular}[c]{@{}l@{}}sDSC\\    \\ (US)\end{tabular} & \multicolumn{1}{l|}{-} & \multicolumn{1}{l|}{0.80 ±   0.07} & \multicolumn{1}{l|}{0.78 ±   0.09} & \multicolumn{1}{l|}{0.7 ±   0.1} & \multicolumn{1}{l|}{0.5 ±   0.1} & - &  \\ \cline{4-10}
				&  &  & \begin{tabular}[c]{@{}l@{}}HD95   (mm)\\    \\ (US)\end{tabular} & \multicolumn{1}{l|}{-} & \multicolumn{1}{l|}{4 ± 5} & \multicolumn{1}{l|}{7 ± 5} & \multicolumn{1}{l|}{20 ±   15} & \multicolumn{1}{l|}{20 ±   15} & - &  \\ \cline{4-10}
				&  &  & \begin{tabular}[c]{@{}l@{}}vDSC\\    \\ (UA)\end{tabular} & \multicolumn{1}{l|}{-} & \multicolumn{1}{l|}{0.93 ±   0.04} & \multicolumn{1}{l|}{0.85 ±   0.06} & \multicolumn{1}{l|}{0.7 ±   0.1} & \multicolumn{1}{l|}{0.7 ±   0.3} & - &  \\ \cline{4-10}
				&  &  & \begin{tabular}[c]{@{}l@{}}sDSC\\    \\ (UA)\end{tabular} & \multicolumn{1}{l|}{-} & \multicolumn{1}{l|}{0.76 ±   0.07} & \multicolumn{1}{l|}{0.70 ±   0.08} & \multicolumn{1}{l|}{0.6 ±   0.1} & \multicolumn{1}{l|}{0.5 ±   0.2} & - &  \\ \cline{4-10}
				&  &  & \begin{tabular}[c]{@{}l@{}}HD95   (mm)\\    \\ (UA)\end{tabular} & \multicolumn{1}{l|}{-} & \multicolumn{1}{l|}{4 ± 2} & \multicolumn{1}{l|}{10 ± 8} & \multicolumn{1}{l|}{20 ±   15} & \multicolumn{1}{l|}{30 ±   30} & - &  \\ \hline
				\multirow{2}{*}{MRI} & \multirow{2}{*}{121} & \multirow{2}{*}{DCT-UNet} & DSC   (T2SPACE) & \multicolumn{1}{l|}{0.70 ±   0.12} & \multicolumn{1}{l|}{0.94 ±   0.10} & \multicolumn{1}{l|}{0.92 ±   0.11} & \multicolumn{1}{l|}{0.84 ± 0.15} & \multicolumn{1}{l|}{-} & - & \multirow{2}{*}{\cite{RN139}} \\ \cline{4-10}
				&  &  & DSC   (T2QM) & \multicolumn{1}{l|}{0.66 ±   0.10} & \multicolumn{1}{l|}{0.98 ±   0.02} & \multicolumn{1}{l|}{0.81 ±   0.04} & \multicolumn{1}{l|}{0.80 ±   0.09} & \multicolumn{1}{l|}{-} & - &  \\ \hline
				\multirow{3}{*}{MRI} & \multirow{3}{*}{195} & \multirow{3}{*}{3D nnU-Net} & DSC & \multicolumn{1}{l|}{-} & \multicolumn{1}{l|}{-} & \multicolumn{1}{l|}{-} & \multicolumn{1}{l|}{-} & \multicolumn{1}{l|}{-} & 0.73   [0.50-0.80] & \multirow{3}{*}{\cite{RN9}} \\ \cline{4-10}
				&  &  & HD95   (mm) & \multicolumn{1}{l|}{-} & \multicolumn{1}{l|}{-} & \multicolumn{1}{l|}{-} & \multicolumn{1}{l|}{-} & \multicolumn{1}{l|}{-} & 6.8   [4.2 –12.5] &  \\ \cline{4-10}
				&  &  & MSD   (mm) & \multicolumn{1}{l|}{-} & \multicolumn{1}{l|}{-} & \multicolumn{1}{l|}{-} & \multicolumn{1}{l|}{-} & \multicolumn{1}{l|}{-} & 1.4   [0.9 – 2.8 &  \\ \hline
				\multirow{2}{*}{MRI} & \multirow{2}{*}{125} & \multirow{2}{*}{3D   U-Net} & DSC & \multicolumn{1}{l|}{0.85 ±   0.03} & \multicolumn{1}{l|}{-} & \multicolumn{1}{l|}{-} & \multicolumn{1}{l|}{-} & \multicolumn{1}{l|}{-} & - & \multirow{2}{*}{\cite{RN142}} \\ \cline{4-10}
				&  &  & HD95   (mm) & \multicolumn{1}{l|}{3.70 ±   0.35} & \multicolumn{1}{l|}{-} & \multicolumn{1}{l|}{-} & \multicolumn{1}{l|}{-} & \multicolumn{1}{l|}{-} & - &  \\ \hline
				\multirow{4}{*}{MRI   and CT} & \multirow{4}{*}{65} & \multirow{4}{*}{Dual-path   CNN} & DSC & \multicolumn{1}{l|}{0.76 ±   0.06} & \multicolumn{1}{l|}{-} & \multicolumn{1}{l|}{-} & \multicolumn{1}{l|}{-} & \multicolumn{1}{l|}{-} & - & \multirow{4}{*}{\cite{RN145}} \\ \cline{4-10}
				&  &  & HD95   (mm) & \multicolumn{1}{l|}{5.99 ±   1.68} & \multicolumn{1}{l|}{-} & \multicolumn{1}{l|}{-} & \multicolumn{1}{l|}{-} & \multicolumn{1}{l|}{-} & - &  \\ \cline{4-10}
				&  &  & Sensitivity & \multicolumn{1}{l|}{0.81 ±   0.04} & \multicolumn{1}{l|}{-} & \multicolumn{1}{l|}{-} & \multicolumn{1}{l|}{-} & \multicolumn{1}{l|}{-} & - &  \\ \cline{4-10}
				&  &  & Precision & \multicolumn{1}{l|}{0.83 ±   0.04} & \multicolumn{1}{l|}{-} & \multicolumn{1}{l|}{-} & \multicolumn{1}{l|}{-} & \multicolumn{1}{l|}{-} & - &  \\ \hline
				\multirow{3}{*}{CT} & \multirow{3}{*}{91} & \multirow{3}{*}{DSD-UNET} & DSC & \multicolumn{1}{l|}{0.83 ±   0.04} & \multicolumn{1}{l|}{0.87 ±   0.03} & \multicolumn{1}{l|}{0.82 ±   0.05} & \multicolumn{1}{l|}{0.65 ±   0.08} & \multicolumn{1}{l|}{0.80 ±   0.06} & - & \multirow{3}{*}{\cite{RN41}} \\ \cline{4-10}
				&  &  & HD   (mm) & \multicolumn{1}{l|}{8.1 ±   2.3} & \multicolumn{1}{l|}{12.1 ±   4.0} & \multicolumn{1}{l|}{9.2 ±   4.6} & \multicolumn{1}{l|}{19.6 ±   8.7} & \multicolumn{1}{l|}{27.8 ±   10.8} & - &  \\ \cline{4-10}
				&  &  & Jaccard   Index & \multicolumn{1}{l|}{0.72 ±   0.04} & \multicolumn{1}{l|}{0.78 ±   0.03} & \multicolumn{1}{l|}{0.72 ±   0.05} & \multicolumn{1}{l|}{0.52 ±   0.08} & \multicolumn{1}{l|}{0.69 ±   0.06} & - &  \\ \hline
				\multirow{3}{*}{CT} & \multirow{3}{*}{200} & \multirow{3}{*}{RefineNet} & DSC & \multicolumn{1}{l|}{0.861} & \multicolumn{1}{l|}{0.86} & \multicolumn{1}{l|}{0.858} & \multicolumn{1}{l|}{0.664} & \multicolumn{1}{l|}{0.563} & - & \multirow{3}{*}{\cite{RN39}} \\ \cline{4-10}
				&  &  & HD   (mm) & \multicolumn{1}{l|}{6.005} & \multicolumn{1}{l|}{19.98} & \multicolumn{1}{l|}{12.27} & \multicolumn{1}{l|}{98.41} & \multicolumn{1}{l|}{68.12} & - &  \\ \cline{4-10}
				&  &  & Overlap   Index & \multicolumn{1}{l|}{0.839} & \multicolumn{1}{l|}{0.783} & \multicolumn{1}{l|}{0.894} & \multicolumn{1}{l|}{0.601} & \multicolumn{1}{l|}{0.811} & - &  \\ \hline
				\multirow{3}{*}{CT} & \multirow{3}{*}{62} & \multirow{3}{*}{nnU-Net} & DSC & \multicolumn{1}{l|}{0.84 ±   0.07} & \multicolumn{1}{l|}{0.94 ±   0.05} & \multicolumn{1}{l|}{0.83 ±   0.07} & \multicolumn{1}{l|}{-} & \multicolumn{1}{l|}{-} & - & \multirow{3}{*}{\cite{RN6}} \\ \cline{4-10}
				&  &  & HD95   (mm) & \multicolumn{1}{l|}{7.42 ±   5.02} & \multicolumn{1}{l|}{3.50 ±   1.96} & \multicolumn{1}{l|}{7.58 ±   5.86} & \multicolumn{1}{l|}{-} & \multicolumn{1}{l|}{-} & - &  \\ \cline{4-10}
				&  &  & MSD   (mm) & \multicolumn{1}{l|}{2.09 ±   1.31} & \multicolumn{1}{l|}{0.94 ±   0.50} & \multicolumn{1}{l|}{3.60 ±   3.49} & \multicolumn{1}{l|}{-} & \multicolumn{1}{l|}{-} & - &  \\ \hline
				\multirow{4}{*}{CT} & \multirow{4}{*}{100} & \multirow{4}{*}{nnU-Net} & DSC & \multicolumn{1}{l|}{0.81 ±   0.05} & \multicolumn{1}{l|}{0.92 ±   0.04} & \multicolumn{1}{l|}{0.84 ±   0.04} & \multicolumn{1}{l|}{-} & \multicolumn{1}{l|}{-} & - & \multirow{4}{*}{\cite{RN5}} \\ \cline{4-10}
				&  &  & HD95   (mm) & \multicolumn{1}{l|}{6.03 ±   2.01} & \multicolumn{1}{l|}{3.00 ±   1.09} & \multicolumn{1}{l|}{5.25 ±   1.78} & \multicolumn{1}{l|}{-} & \multicolumn{1}{l|}{-} & - &  \\ \cline{4-10}
				&  &  & MSD   (mm) & \multicolumn{1}{l|}{2.23 ±   0.75} & \multicolumn{1}{l|}{0.84 ±   0.30} & \multicolumn{1}{l|}{1.36 ±   0.43} & \multicolumn{1}{l|}{-} & \multicolumn{1}{l|}{-} & - &  \\ \cline{4-10}
				&  &  & Precision & \multicolumn{1}{l|}{0.80} & \multicolumn{1}{l|}{0.91} & \multicolumn{1}{l|}{0.84} & \multicolumn{1}{l|}{-} & \multicolumn{1}{l|}{-} & - &  \\ \hline
				\multirow{2}{*}{CT} & \multirow{2}{*}{60} & \multirow{2}{*}{Modified   CNN} & DSC & \multicolumn{1}{l|}{0.87} & \multicolumn{1}{l|}{0.94} & \multicolumn{1}{l|}{0.86} & \multicolumn{1}{l|}{0.79} & \multicolumn{1}{l|}{0.92} & - & \multirow{2}{*}{\cite{RN40}} \\ \cline{4-10}
				&  &  & HD95   (mm) & \multicolumn{1}{l|}{1.45} & \multicolumn{1}{l|}{4.52} & \multicolumn{1}{l|}{2.52} & \multicolumn{1}{l|}{10.92} & \multicolumn{1}{l|}{8.83} & - &  \\ \hline
				\multirow{2}{*}{CT} & \multirow{2}{*}{98} & \multirow{2}{*}{SEResU-Net} & DSC & \multicolumn{1}{l|}{0.81 ±   0.05} & \multicolumn{1}{l|}{0.92 ±   0.03} & \multicolumn{1}{l|}{0.85 ±   0.05} & \multicolumn{1}{l|}{0.60 ±   0.12} & \multicolumn{1}{l|}{0.83 ±   0.09} & - & \multirow{2}{*}{\cite{RN135}} \\ \cline{4-10}
				&  &  & HD95   (mm) & \multicolumn{1}{l|}{5.23 ±   1.39} & \multicolumn{1}{l|}{4.75 ±   1.48} & \multicolumn{1}{l|}{4.06 ±   1.68} & \multicolumn{1}{l|}{30 ±   8.16} & \multicolumn{1}{l|}{20.5 ±   9.88} & - &  \\ \hline
				\multirow{3}{*}{CT} & \multirow{3}{*}{321} & \multirow{3}{*}{Prompt-nnUnet} & DSC & \multicolumn{1}{l|}{0.96 ±   0.02} & \multicolumn{1}{l|}{-} & \multicolumn{1}{l|}{0.91 ±   0.02} & \multicolumn{1}{l|}{-} & \multicolumn{1}{l|}{-} & - & \multirow{3}{*}{\cite{RN137}} \\ \cline{4-10}
				&  &  & HD95   (mm) & \multicolumn{1}{l|}{1.66 ±   1.11} & \multicolumn{1}{l|}{-} & \multicolumn{1}{l|}{3.07 ±   0.94} & \multicolumn{1}{l|}{-} & \multicolumn{1}{l|}{-} & - &  \\ \cline{4-10}
				&  &  & IoU & \multicolumn{1}{l|}{0.92 ±   0.04} & \multicolumn{1}{l|}{-} & \multicolumn{1}{l|}{0.84 ±   0.03} & \multicolumn{1}{l|}{-} & \multicolumn{1}{l|}{-} & - &  \\ \hline
				\multirow{3}{*}{CT} & \multirow{3}{*}{113} & \multirow{3}{*}{ResU-Net} & DSC & \multicolumn{1}{l|}{-} & \multicolumn{1}{l|}{0.96 ±   0.04} & \multicolumn{1}{l|}{0.97 ±   0.02} & \multicolumn{1}{l|}{0.92 ±   0.03} & \multicolumn{1}{l|}{-} & - & \multirow{3}{*}{\cite{RN2}} \\ \cline{4-10}
				&  &  & HD   (mm) & \multicolumn{1}{l|}{-} & \multicolumn{1}{l|}{4.05 ±   5.17} & \multicolumn{1}{l|}{1.96 ±   2.19} & \multicolumn{1}{l|}{3.15 ±   2.03} & \multicolumn{1}{l|}{-} & - &  \\ \cline{4-10}
				&  &  & MSD   (mm) & \multicolumn{1}{l|}{-} & \multicolumn{1}{l|}{1.04 ±   0.97} & \multicolumn{1}{l|}{0.45 ±   0.09} & \multicolumn{1}{l|}{0.79 ±   0.25} & \multicolumn{1}{l|}{-} & - &  \\ \hline
				CT & 51 & 3D   U-Net and Long Short-Term Memory & DSC & \multicolumn{1}{l|}{0.87 ±   0.063} & \multicolumn{1}{l|}{0.86 ±   0.049} & \multicolumn{1}{l|}{0.77 ±   0.084} & \multicolumn{1}{l|}{0.73 ±   0.0102} & \multicolumn{1}{l|}{-} & 0.72 ±   0.091 & \cite{RN8} \\ \hline
				\multirow{5}{*}{CT} & \multirow{5}{*}{53} & \multirow{5}{*}{Mnet\_IM} & sDSC & \multicolumn{1}{l|}{0.81 ±   0.007} & \multicolumn{1}{l|}{-} & \multicolumn{1}{l|}{-} & \multicolumn{1}{l|}{-} & \multicolumn{1}{l|}{-} & - & \multirow{5}{*}{\cite{RN147}} \\ \cline{4-10}
				&  &  & vDSC & \multicolumn{1}{l|}{0.88 ±   0.001} & \multicolumn{1}{l|}{-} & \multicolumn{1}{l|}{-} & \multicolumn{1}{l|}{-} & \multicolumn{1}{l|}{-} & - &  \\ \cline{4-10}
				&  &  & Surface   Overlap & \multicolumn{1}{l|}{0.78 ±   0.007} & \multicolumn{1}{l|}{-} & \multicolumn{1}{l|}{-} & \multicolumn{1}{l|}{-} & \multicolumn{1}{l|}{-} & - &  \\ \cline{4-10}
				&  &  & HD   (mm) & \multicolumn{1}{l|}{3.20 ±   2.00} & \multicolumn{1}{l|}{-} & \multicolumn{1}{l|}{-} & \multicolumn{1}{l|}{-} & \multicolumn{1}{l|}{-} & - &  \\ \cline{4-10}
				&  &  & MSD   (mm) & \multicolumn{1}{l|}{0.69 ±   0.06} & \multicolumn{1}{l|}{-} & \multicolumn{1}{l|}{-} & \multicolumn{1}{l|}{-} & \multicolumn{1}{l|}{-} & - &  \\ \hline
			\end{tabular}%
		}
	Note: Abbreviations: US (unseen scanner), UA (unseen applicator), vDSC (volumetric dice similarity coefficient), sDSC (surface dice similarity coefficient), DSD-UNET (dilated convolution and deep supervision U-Net), SEResU-Net (U-Net with squeeze-and-excitation ResNet), IoU (intersection over union), Mnet\_IM (improved M-Net model).
	\end{table}
	
	\noindent 
	\subsubsection{Prostate}
	
	For prostate brachytherapy, the target is typically visualized using PET/CT, TRUS, MRI, or CT during different procedures. Several studies have employed deep learning to segment the prostate, lesion, and OARs across different imaging modalities. 
	
	Dose boost on the dominant intraprostatic lesion (DIL) could potentially enhance the treatment outcomes \cite{RN118, RN116, RN117}. Accurately segmenting the DIL is thus important for an effective DIL boost in prostate HDR brachytherapy plans, to ensure precise delivery of boost dose. PET/CT imaging can provide detailed morphological/anatomical information about the prostate and DIL. Matkovic et al. \cite{RN1} used the Cascaded Regional-Net to automatically segment the prostate and DIL on PET/CT images. The Cascaded Regional-Net used a Dual Attention Network to extract deep features and identify the volume-of-interest (VOI) of the prostate, narrowing the location range of the DIL. Subsequently, a mask scoring regional convolutional neural network (MSR-CNN) detected the VOIs of the DILs and segmented the DIL from the prostate VOI. The MSDs were 0.666 ± 0.696 mm and 0.814 ± 1.002 mm, with DSCs of 0.932 ± 0.059 and 0.801 ± 0.178 for the prostate and DIL, respectively. The DSC for DIL is relatively low due to its small size and irregular shape. The CT scans provide additional anatomical structures of the patient but may add complexity in lesion segmentation. Wang et al. \cite{RN10} conducted a similar study using Cascaded U-net to segment the lesions on PET scans with and without CT information. More lesions were detected on PET only compared to PET/CT scans (153/155 vs. 144/155), but there was no statistically significant difference in the DSC and HD95 between the PET only and PET/CT images as indicated in Table 3. Additionally, other studies investigated lesion segmentation using different methods. Li et al. \cite{RN43} tested their model on 56 PET/CT scans from external institutions and showed no statistically significant difference compared to the internal testing results. 
	
	Prostate brachytherapy relies on TRUS images to guide implants. Accurate delineation of the prostate and OARs may enable a TRUS-based planning workflow, eliminating the need for additional CT or MR scans. However, as illustrated in Figure 5, the segmentation process is challenging due to the unclear boundary between the prostate and rectum, making it highly dependent on the clinician’ experience.
	
	\begin{figure}
		\centering
		\noindent \includegraphics*[width=6.50in, height=4.20in, keepaspectratio=true]{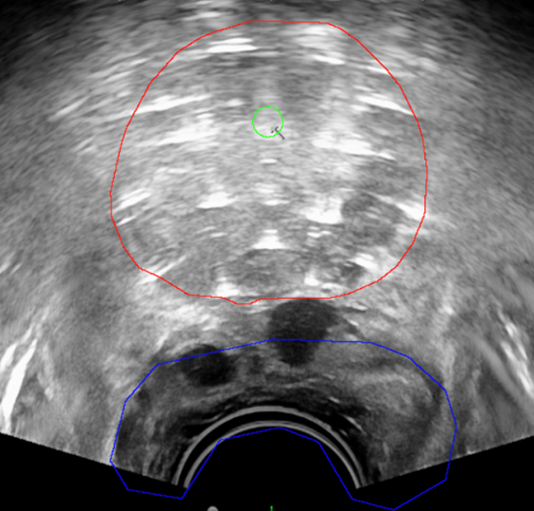}
		
		\noindent Figure 5: Example of organ segmentation on TRUS used in prostate brachytherapy. Red: prostate; blue: rectum; green: urethra.
	\end{figure}

	Several studies have successfully used different deep-learning models to segment the prostate on TRUS images, and the detailed results are summarized in Table 3 \cite{RN47, RN44, RN52, RN50, RN46, RN45, RN48, RN13, RN49, RN51}. Among these studies, the most accurate contours are reported from the semi-automatic models by Girum et al. \cite{RN52} and Peng et al. \cite{RN13}, both of which achieved DSCs of more than 0.96 for the prostate segmentation. Girum et al. used a weakly supervised deep learning method, which is a fully connected CNN with a prior knowledge generator block and a segmentation block. The weakly supervised method is more effective and less time-consuming than training with fully labeled image data. The segmentation highly accurate DSC of 0.969 ± 0.009 and HD of 4.25 ± 4.58 mm on TRUS. Interestingly, this model can also be directly implemented on CT, despite being trained on TRUS images, with only slightly inferior results (DSC = 0.954 ± 0.009 and HD = 5.17 ± 1.41 mm). This study showed the potential for model adaptation between different imaging modalities, which enables the model to be implemented in imaging modalities with limited data. Peng et al. built an A-ProSeg model, which combines a closed-principle-curve-based method, a global closed polygonal segment method, and memory-based differential evolution model to identify prostate vertices and create a smooth prostate contour. The model demonstrated robustness as was trained and tested on a diverse dataset containing 226 patients with 945 TRUS slices in total. It achieved a DSC of 0.962 ± 0.024 and a HD of 1.9 ± 0.9 mm and maintained a similar segmentation performance even when the TRUS images were set to a low signal-to-noise ratio (SNR) of 0.8. It is worth to note that both methods were semi-automatic, requiring pseudo-landmarks or radiologist-defined points as inputs, which may be prone to user errors or inter-observer variability.
	
	Fully automatic models can further simplify the contouring process as they do not require human intervention. Orlando et al. \cite{RN11} trained and validated an accurate fully automatic model for segmenting the prostate on TRUS on a large dataset of 246 patients. A 2D modified U-Net was built to predict the 2D radial slices from TRUS images and the predicted prostate slices were reconstructed in 3D. Using this configuration, the median DSC was 0.94, HD was 2.89 mm, volume percent differences (VPD) was 5.78\%, and MSD was 0.89 mm. This group conducted another study comparing U-Net and U-Net++ with different training configurations. \cite{RN12} The U-Net++ model achieved the most optimal segmentation performance with a training dataset size of 1,000 2D images, regardless of the imaging acquisition type (end-fire, side-fire, or mixed). Interestingly, increasing the training data size does not significantly enhance the segmentation results but add additional training time.
	
	In prostate brachytherapy, the OARs include the rectum, bladder, and seminal vesicles due to their proximity to the prostate. An efficient and accurate segmentation method of these organs is essential to minimize unnecessary radiation dose and reduce treatment toxicity. Three groups have provided methods to segment the prostate and OARs on TRUS, MRI, or CT images.
	
	Lei et al. \cite{RN148} designed an anchor-free mask CNN that utilized a fully convolutional one-state object detector that simultaneously segment the prostate, bladder, rectum, and urethra on 3D TRUS images within 5 seconds per patient. This model provides fast segmentation on all organs but has limitations in the segmentation accuracy of the bladder, constrained by the low contrast on ultrasound images. Sanders et al. \cite{RN14} trained 120 models with different combinations of 18 FCNs and different convolutional encoders, to segment the prostate and the OARs in T1, T2/T1, and T2-weighted MRIs. Among all combinations, an FCN with a DenseNet201 encoder yielded the most optimal results besides the external urinary sphincter and the seminal vesicles, due to their complex shapes.  While the model was trained exclusively on T2-weighted and T2/T1-weighted contrast images, it can also be transferred to T1-weighted MRI with slightly reduced accuracy. Duan et al. \cite{RN16} conducted a study to assess the efficacy of the commercial deep learning auto-segmentation (DLAS) software in automatically segmenting the male pelvis on CT scans. Retraining the DLAS software with institutional data, the segmentation results improved for the prostate and OARs compared to the built-in setup.
	
	\begin{table}[]
		\caption{}
		\label{tab:my-table}
		\resizebox{\textwidth}{!}{%
			\begin{tabular}{|l|l|l|l|llllll|l|}
				\hline
				\multirow{2}{*}{Image   Modality} & \multirow{2}{*}{Number   of Patients} & \multirow{2}{*}{Model} & \multirow{2}{*}{Evaluation   Parameters} & \multicolumn{6}{l|}{Results} & \multirow{2}{*}{Citation} \\ \cline{5-10}
				&  &  &  & \multicolumn{1}{l|}{Prostate} & \multicolumn{1}{l|}{Prostate   Lesion} & \multicolumn{1}{l|}{Bladder} & \multicolumn{1}{l|}{Rectum} & \multicolumn{1}{l|}{Urethra} & Seminal   Vesicles &  \\ \hline
				\multirow{2}{*}{PET/CT} & \multirow{2}{*}{49} & \multirow{2}{*}{Cascaded   Regional Net and MSR-CNN} & DSC & \multicolumn{1}{l|}{0.932   ± 0.059} & \multicolumn{1}{l|}{0.801   ± 0.178} & \multicolumn{1}{l|}{-} & \multicolumn{1}{l|}{-} & \multicolumn{1}{l|}{-} & - & \multirow{2}{*}{\cite{RN1}} \\ \cline{4-10}
				&  &  & MSD   (mm) & \multicolumn{1}{l|}{0.666   ± 0.696} & \multicolumn{1}{l|}{0.814   ± 1.002} & \multicolumn{1}{l|}{-} & \multicolumn{1}{l|}{-} & \multicolumn{1}{l|}{-} & - &  \\ \hline
				\multirow{6}{*}{PET/CT} & \multirow{6}{*}{84} & \multirow{6}{*}{U-net,   Cascaded U-net, and cascaded detection segmentation network} & DSC   (PET/CT) & \multicolumn{1}{l|}{-} & \multicolumn{1}{l|}{0.68 ±   0.15} & \multicolumn{1}{l|}{-} & \multicolumn{1}{l|}{-} & \multicolumn{1}{l|}{-} & - & \multirow{6}{*}{\cite{RN10}} \\ \cline{4-10}
				&  &  & DSC   (PET only) & \multicolumn{1}{l|}{-} & \multicolumn{1}{l|}{0.68 ±   0.17} & \multicolumn{1}{l|}{-} & \multicolumn{1}{l|}{-} & \multicolumn{1}{l|}{-} & - &  \\ \cline{4-10}
				&  &  & HD95   (mm) (PET/CT) & \multicolumn{1}{l|}{-} & \multicolumn{1}{l|}{3.98 ±   2.23} & \multicolumn{1}{l|}{-} & \multicolumn{1}{l|}{-} & \multicolumn{1}{l|}{-} & - &  \\ \cline{4-10}
				&  &  & HD95   (mm) (PET only) & \multicolumn{1}{l|}{-} & \multicolumn{1}{l|}{4.16 ±   2.33} & \multicolumn{1}{l|}{-} & \multicolumn{1}{l|}{-} & \multicolumn{1}{l|}{-} & - &  \\ \cline{4-10}
				&  &  & Detection   rate (PET/CT) & \multicolumn{1}{l|}{-} & \multicolumn{1}{l|}{144/155} & \multicolumn{1}{l|}{-} & \multicolumn{1}{l|}{-} & \multicolumn{1}{l|}{-} & - &  \\ \cline{4-10}
				&  &  & Detection   rate (PET only) & \multicolumn{1}{l|}{-} & \multicolumn{1}{l|}{153/155} & \multicolumn{1}{l|}{-} & \multicolumn{1}{l|}{-} & \multicolumn{1}{l|}{-} & - &  \\ \hline
				\multirow{8}{*}{PET/CT} & \multirow{8}{*}{137} & \multirow{8}{*}{UNETR} & DSC   (internal) & \multicolumn{1}{l|}{-} & \multicolumn{1}{l|}{0.70} & \multicolumn{1}{l|}{-} & \multicolumn{1}{l|}{-} & \multicolumn{1}{l|}{-} & - & \multirow{8}{*}{\cite{RN43}} \\ \cline{4-10}
				&  &  & DSC   (external) & \multicolumn{1}{l|}{-} & \multicolumn{1}{l|}{0.68} & \multicolumn{1}{l|}{-} & \multicolumn{1}{l|}{-} & \multicolumn{1}{l|}{-} & - &  \\ \cline{4-10}
				&  &  & IoU (internal) & \multicolumn{1}{l|}{-} & \multicolumn{1}{l|}{0.566} & \multicolumn{1}{l|}{-} & \multicolumn{1}{l|}{-} & \multicolumn{1}{l|}{-} & - &  \\ \cline{4-10}
				&  &  & IoU   (external) & \multicolumn{1}{l|}{-} & \multicolumn{1}{l|}{0.548} & \multicolumn{1}{l|}{-} & \multicolumn{1}{l|}{-} & \multicolumn{1}{l|}{-} & - &  \\ \cline{4-10}
				&  &  & Precision   (internal) & \multicolumn{1}{l|}{-} & \multicolumn{1}{l|}{0.809} & \multicolumn{1}{l|}{-} & \multicolumn{1}{l|}{-} & \multicolumn{1}{l|}{-} & - &  \\ \cline{4-10}
				&  &  & Precision   (external) & \multicolumn{1}{l|}{-} & \multicolumn{1}{l|}{0.749} & \multicolumn{1}{l|}{-} & \multicolumn{1}{l|}{-} & \multicolumn{1}{l|}{-} & - &  \\ \cline{4-10}
				&  &  & Recall   (internal) & \multicolumn{1}{l|}{-} & \multicolumn{1}{l|}{0.66} & \multicolumn{1}{l|}{-} & \multicolumn{1}{l|}{-} & \multicolumn{1}{l|}{-} & - &  \\ \cline{4-10}
				&  &  & Recall   (external) & \multicolumn{1}{l|}{-} & \multicolumn{1}{l|}{0.74} & \multicolumn{1}{l|}{-} & \multicolumn{1}{l|}{-} & \multicolumn{1}{l|}{-} & - &  \\ \hline
				\multirow{6}{*}{TRUS   and CT} & \multirow{6}{*}{145} & \multirow{6}{*}{Weakly-supervised   CNN and unsupervised CNN} & DSC   (TRUS) & \multicolumn{1}{l|}{0.969   ± 0.009} & \multicolumn{1}{l|}{-} & \multicolumn{1}{l|}{-} & \multicolumn{1}{l|}{-} & \multicolumn{1}{l|}{-} & - & \multirow{6}{*}{\cite{RN52}} \\ \cline{4-10}
				&  &  & DSC   (CT) & \multicolumn{1}{l|}{0.954   ± 0.009} & \multicolumn{1}{l|}{-} & \multicolumn{1}{l|}{-} & \multicolumn{1}{l|}{-} & \multicolumn{1}{l|}{-} & - &  \\ \cline{4-10}
				&  &  & 3D HD   (mm) (TRUS) & \multicolumn{1}{l|}{4.25 ±   4.58} & \multicolumn{1}{l|}{-} & \multicolumn{1}{l|}{-} & \multicolumn{1}{l|}{-} & \multicolumn{1}{l|}{-} & - &  \\ \cline{4-10}
				&  &  & 3D HD   (mm) (CT) & \multicolumn{1}{l|}{5.17 ±   1.41} & \multicolumn{1}{l|}{-} & \multicolumn{1}{l|}{-} & \multicolumn{1}{l|}{-} & \multicolumn{1}{l|}{-} & - &  \\ \cline{4-10}
				&  &  & Volumetric   Overlap Ratio (TRUS) & \multicolumn{1}{l|}{0.939   ± 0.180} & \multicolumn{1}{l|}{-} & \multicolumn{1}{l|}{-} & \multicolumn{1}{l|}{-} & \multicolumn{1}{l|}{-} & - &  \\ \cline{4-10}
				&  &  & Volumetric   Overlap Ratio (CT) & \multicolumn{1}{l|}{0.913   ± 0.170} & \multicolumn{1}{l|}{-} & \multicolumn{1}{l|}{-} & \multicolumn{1}{l|}{-} & \multicolumn{1}{l|}{-} & - &  \\ \hline
				\multirow{3}{*}{TRUS} & \multirow{3}{*}{226} & \multirow{3}{*}{A-ProSeg} & DSC & \multicolumn{1}{l|}{0.962   ± 0.024} & \multicolumn{1}{l|}{-} & \multicolumn{1}{l|}{-} & \multicolumn{1}{l|}{-} & \multicolumn{1}{l|}{-} & - & \multirow{3}{*}{\cite{RN13}} \\ \cline{4-10}
				&  &  & JI & \multicolumn{1}{l|}{0.944   ± 0.033} & \multicolumn{1}{l|}{-} & \multicolumn{1}{l|}{-} & \multicolumn{1}{l|}{-} & \multicolumn{1}{l|}{-} & - &  \\ \cline{4-10}
				&  &  & Accuracy & \multicolumn{1}{l|}{95.7 ±   2.7\%} & \multicolumn{1}{l|}{-} & \multicolumn{1}{l|}{-} & \multicolumn{1}{l|}{-} & \multicolumn{1}{l|}{-} & - &  \\ \hline
				\multirow{6}{*}{TRUS} & \multirow{6}{*}{246} & \multirow{6}{*}{2D   modified U-Net and 3D reconstruction} & DSC & \multicolumn{1}{l|}{0.941   [0.926, 0.949]} & \multicolumn{1}{l|}{-} & \multicolumn{1}{l|}{-} & \multicolumn{1}{l|}{-} & \multicolumn{1}{l|}{-} & - & \multirow{6}{*}{\cite{RN11}} \\ \cline{4-10}
				&  &  & Precision & \multicolumn{1}{l|}{93.2   [88.8, 95.4] \%} & \multicolumn{1}{l|}{-} & \multicolumn{1}{l|}{-} & \multicolumn{1}{l|}{-} & \multicolumn{1}{l|}{-} & - &  \\ \cline{4-10}
				&  &  & Recall & \multicolumn{1}{l|}{96.0   [93.1, 98.5] \%} & \multicolumn{1}{l|}{-} & \multicolumn{1}{l|}{-} & \multicolumn{1}{l|}{-} & \multicolumn{1}{l|}{-} & - &  \\ \cline{4-10}
				&  &  & VPD & \multicolumn{1}{l|}{5.78   [2.49, 11.50] \%} & \multicolumn{1}{l|}{-} & \multicolumn{1}{l|}{-} & \multicolumn{1}{l|}{-} & \multicolumn{1}{l|}{-} & - &  \\ \cline{4-10}
				&  &  & HD   (mm) & \multicolumn{1}{l|}{2.89   [2.37, 4.35]} & \multicolumn{1}{l|}{-} & \multicolumn{1}{l|}{-} & \multicolumn{1}{l|}{-} & \multicolumn{1}{l|}{-} & - &  \\ \cline{4-10}
				&  &  & MSD   (mm) & \multicolumn{1}{l|}{0.89   [0.73, 1.09]} & \multicolumn{1}{l|}{-} & \multicolumn{1}{l|}{-} & \multicolumn{1}{l|}{-} & \multicolumn{1}{l|}{-} & - &  \\ \hline
				\multirow{3}{*}{TRUS} & \multirow{3}{*}{44} & \multirow{3}{*}{Multidirectional   Deeply Supervised V-Net} & DSC & \multicolumn{1}{l|}{0.92 ±   0.03} & \multicolumn{1}{l|}{-} & \multicolumn{1}{l|}{-} & \multicolumn{1}{l|}{-} & \multicolumn{1}{l|}{-} & - & \multirow{3}{*}{\cite{RN45}} \\ \cline{4-10}
				&  &  & HD   (mm) & \multicolumn{1}{l|}{3.94 ±   1.55} & \multicolumn{1}{l|}{-} & \multicolumn{1}{l|}{-} & \multicolumn{1}{l|}{-} & \multicolumn{1}{l|}{-} & - &  \\ \cline{4-10}
				&  &  & MSD   (mm) & \multicolumn{1}{l|}{0.60 ±   0.23} & \multicolumn{1}{l|}{-} & \multicolumn{1}{l|}{-} & \multicolumn{1}{l|}{-} & \multicolumn{1}{l|}{-} & - &  \\ \hline
				\multirow{2}{*}{TRUS} & \multirow{2}{*}{675} & \multirow{2}{*}{CNN} & DSC & \multicolumn{1}{l|}{0.939   ± 0.035} & \multicolumn{1}{l|}{-} & \multicolumn{1}{l|}{-} & \multicolumn{1}{l|}{-} & \multicolumn{1}{l|}{-} & - & \multirow{2}{*}{\cite{RN46}} \\ \cline{4-10}
				&  &  & HD   (mm) & \multicolumn{1}{l|}{2.7 ±   2.3} & \multicolumn{1}{l|}{-} & \multicolumn{1}{l|}{-} & \multicolumn{1}{l|}{-} & \multicolumn{1}{l|}{-} & - &  \\ \hline
				\multirow{4}{*}{TRUS} & \multirow{4}{*}{145} & \multirow{4}{*}{Encoder–decoder   CNN and DNN} & DSC & \multicolumn{1}{l|}{0.88 ±   0.02} & \multicolumn{1}{l|}{-} & \multicolumn{1}{l|}{-} & \multicolumn{1}{l|}{-} & \multicolumn{1}{l|}{-} & - & \multirow{4}{*}{\cite{RN44}} \\ \cline{4-10}
				&  &  & HD95   (mm) & \multicolumn{1}{l|}{2.01 ±   0.54} & \multicolumn{1}{l|}{-} & \multicolumn{1}{l|}{-} & \multicolumn{1}{l|}{-} & \multicolumn{1}{l|}{-} & - &  \\ \cline{4-10}
				&  &  & Accuracy & \multicolumn{1}{l|}{96 ±   1\%} & \multicolumn{1}{l|}{-} & \multicolumn{1}{l|}{-} & \multicolumn{1}{l|}{-} & \multicolumn{1}{l|}{-} & - &  \\ \cline{4-10}
				&  &  & MSD   (mm) & \multicolumn{1}{l|}{0.1 ±   0.06} & \multicolumn{1}{l|}{-} & \multicolumn{1}{l|}{-} & \multicolumn{1}{l|}{-} & \multicolumn{1}{l|}{-} & - &  \\ \hline
				\multirow{3}{*}{TRUS} & \multirow{3}{*}{598} & \multirow{3}{*}{ResU-Net} & DSC & \multicolumn{1}{l|}{0.937   ± 0.037} & \multicolumn{1}{l|}{-} & \multicolumn{1}{l|}{-} & \multicolumn{1}{l|}{-} & \multicolumn{1}{l|}{-} & - & \multirow{3}{*}{\cite{RN47}} \\ \cline{4-10}
				&  &  & HD   (mm) & \multicolumn{1}{l|}{3.0 ±   2.05} & \multicolumn{1}{l|}{-} & \multicolumn{1}{l|}{-} & \multicolumn{1}{l|}{-} & \multicolumn{1}{l|}{-} & - &  \\ \cline{4-10}
				&  &  & MSD   (mm) & \multicolumn{1}{l|}{1.05 ±   0.71} & \multicolumn{1}{l|}{-} & \multicolumn{1}{l|}{-} & \multicolumn{1}{l|}{-} & \multicolumn{1}{l|}{-} & - &  \\ \hline
				\multirow{6}{*}{TRUS} & \multirow{6}{*}{590} & \multirow{6}{*}{Multi-label   method with K-SVD} & Volumetric   Error (CTV) & \multicolumn{1}{l|}{9.95 ±   3.53\%} & \multicolumn{1}{l|}{-} & \multicolumn{1}{l|}{-} & \multicolumn{1}{l|}{-} & \multicolumn{1}{l|}{-} & - & \multirow{6}{*}{\cite{RN48}} \\ \cline{4-10}
				&  &  & Volumetric   Error (PTV) & \multicolumn{1}{l|}{8.84 ±   3.13\%} & \multicolumn{1}{l|}{-} & \multicolumn{1}{l|}{-} & \multicolumn{1}{l|}{-} & \multicolumn{1}{l|}{-} & - &  \\ \cline{4-10}
				&  &  & HD   (mm) (CTV) & \multicolumn{1}{l|}{5.40 ±   1.38} & \multicolumn{1}{l|}{-} & \multicolumn{1}{l|}{-} & \multicolumn{1}{l|}{-} & \multicolumn{1}{l|}{-} & - &  \\ \cline{4-10}
				&  &  & HD   (mm) (PTV) & \multicolumn{1}{l|}{5.48 ±   1.51} & \multicolumn{1}{l|}{-} & \multicolumn{1}{l|}{-} & \multicolumn{1}{l|}{-} & \multicolumn{1}{l|}{-} & - &  \\ \cline{4-10}
				&  &  & MSD   (mm) (CTV) & \multicolumn{1}{l|}{0.98 ±   0.39} & \multicolumn{1}{l|}{-} & \multicolumn{1}{l|}{-} & \multicolumn{1}{l|}{-} & \multicolumn{1}{l|}{-} & - &  \\ \cline{4-10}
				&  &  & MSD   (mm) (PTV) & \multicolumn{1}{l|}{1.19 ±   0.48} & \multicolumn{1}{l|}{-} & \multicolumn{1}{l|}{-} & \multicolumn{1}{l|}{-} & \multicolumn{1}{l|}{-} & - &  \\ \hline
				\multirow{3}{*}{TRUS} & \multirow{3}{*}{315} & \multirow{3}{*}{PTN   and CPTTA} & DSC & \multicolumn{1}{l|}{0.899 ±   0.035} & \multicolumn{1}{l|}{-} & \multicolumn{1}{l|}{-} & \multicolumn{1}{l|}{-} & \multicolumn{1}{l|}{-} & - & \multirow{3}{*}{\cite{RN49}} \\ \cline{4-10}
				&  &  & HD   (mm) & \multicolumn{1}{l|}{7.07 ±   3.19} & \multicolumn{1}{l|}{-} & \multicolumn{1}{l|}{-} & \multicolumn{1}{l|}{-} & \multicolumn{1}{l|}{-} & - &  \\ \cline{4-10}
				&  &  & MSD   (mm) & \multicolumn{1}{l|}{1.30 ±   0.61} & \multicolumn{1}{l|}{-} & \multicolumn{1}{l|}{-} & \multicolumn{1}{l|}{-} & \multicolumn{1}{l|}{-} & - &  \\ \hline
				\multirow{3}{*}{TRUS} & \multirow{3}{*}{132} & \multirow{3}{*}{2D   U-Net CNN} & DSC & \multicolumn{1}{l|}{0.872   [0.841, 0.888]} & \multicolumn{1}{l|}{-} & \multicolumn{1}{l|}{-} & \multicolumn{1}{l|}{-} & \multicolumn{1}{l|}{-} & - & \multirow{3}{*}{\cite{RN50}} \\ \cline{4-10}
				&  &  & HD   (mm) & \multicolumn{1}{l|}{6.0   [5.3, 8.0]} & \multicolumn{1}{l|}{-} & \multicolumn{1}{l|}{-} & \multicolumn{1}{l|}{-} & \multicolumn{1}{l|}{-} & - &  \\ \cline{4-10}
				&  &  & MSD   (mm) & \multicolumn{1}{l|}{1.6   [1.2, 2.0]} & \multicolumn{1}{l|}{-} & \multicolumn{1}{l|}{-} & \multicolumn{1}{l|}{-} & \multicolumn{1}{l|}{-} & - &  \\ \hline
				TRUS   and MRI & 598 & End-to-end   CNN & DSC & \multicolumn{1}{l|}{0.909   ± 0.022} & \multicolumn{1}{l|}{-} & \multicolumn{1}{l|}{-} & \multicolumn{1}{l|}{-} & \multicolumn{1}{l|}{-} & - & \cite{RN51} \\ \hline
				\multirow{4}{*}{TRUS} & \multirow{4}{*}{83} & \multirow{4}{*}{Anchor-free   mask CNN} & DSC   (cross validation) & \multicolumn{1}{l|}{0.93 ±   0.03} & \multicolumn{1}{l|}{-} & \multicolumn{1}{l|}{0.75 ±   0.012} & \multicolumn{1}{l|}{0.90 ±   0.07} & \multicolumn{1}{l|}{0.86 ±   0.07} & - & \multirow{4}{*}{\cite{RN148}} \\ \cline{4-10}
				&  &  & DSC   (hold-out) & \multicolumn{1}{l|}{0.94 ±   0.03} & \multicolumn{1}{l|}{-} & \multicolumn{1}{l|}{0.76 ±   0.13} & \multicolumn{1}{l|}{0.92 ±   0.03} & \multicolumn{1}{l|}{0.85 ±   0.06} & - &  \\ \cline{4-10}
				&  &  & HD   (mm) (cross validation) & \multicolumn{1}{l|}{2.28 ±   0.64} & \multicolumn{1}{l|}{-} & \multicolumn{1}{l|}{2.58 ±   0.7} & \multicolumn{1}{l|}{1.65 ±   0.52} & \multicolumn{1}{l|}{1.85 ±   1.71} & - &  \\ \cline{4-10}
				&  &  & HD   (mm) (hold-out) & \multicolumn{1}{l|}{2.27 ±   0.79} & \multicolumn{1}{l|}{-} & \multicolumn{1}{l|}{2.93 ±   1.29} & \multicolumn{1}{l|}{1.90 ±   0.28} & \multicolumn{1}{l|}{1.81 ±   0.72} & - &  \\ \hline
				\multirow{2}{*}{MRI} & \multirow{2}{*}{200} & \multirow{2}{*}{2D and   3D U-Net FCNs} & DSC   (T2) & \multicolumn{1}{l|}{0.90 ±   0.04} & \multicolumn{1}{l|}{-} & \multicolumn{1}{l|}{0.96 ±   0.04} & \multicolumn{1}{l|}{0.91 ±   0.06} & \multicolumn{1}{l|}{-} & 0.80 ±   0.12 & \multirow{2}{*}{\cite{RN14}} \\ \cline{4-10}
				&  &  & DSC   (T1) & \multicolumn{1}{l|}{0.82 ±   0.07} & \multicolumn{1}{l|}{-} & \multicolumn{1}{l|}{0.88 ±   0.05} & \multicolumn{1}{l|}{0.87 ±   0.06} & \multicolumn{1}{l|}{-} & 0.46 ±   0.21 &  \\ \hline
				\multirow{2}{*}{CT} & \multirow{2}{*}{215} & \multirow{2}{*}{DLAS} & DSC   (re-trained) & \multicolumn{1}{l|}{0.82} & \multicolumn{1}{l|}{-} & \multicolumn{1}{l|}{-} & \multicolumn{1}{l|}{0.92} & \multicolumn{1}{l|}{-} & 0.48 & \multirow{2}{*}{\cite{RN16}} \\ \cline{4-10}
				&  &  & DSC   (built-in) & \multicolumn{1}{l|}{0.73} & \multicolumn{1}{l|}{-} & \multicolumn{1}{l|}{-} & \multicolumn{1}{l|}{0.81} & \multicolumn{1}{l|}{-} & 0.37 &  \\ \hline
			\end{tabular}%
		}
	Note: Abbreviations: DNN (deep neural network), PTN (polar transform network), CPTTA (centroid perturbed test-time augmentation), K-SVD (K-singular value decomposition), A-ProSeg (accurate prostate segmentation framework).
	\end{table}
	
	\noindent 
	\subsection{Other Imaging Applications }
	
	In addition to aiding with image registration and segmentation, AI can enhance medical image quality and generate synthetic images to improve diagnosis and other steps in the brachytherapy workflow. These applications are presented in Table 4. 
	
	Metal artifacts on CT images may complicate organ and applicator visualization. Figure 6 is an example of how metal artifacts caused by hip prothesis can complicate the visualization of patient anatomy and applicators on CT scans. Huang et al. \cite{RN21} built a residual learning method based on CNN (RL-ARCNN) to reduce metal artifacts on CT images for GYN cancer brachytherapy. They generated 600 simulated artifact image slices from 20 GYN cancer patients to train and validate the RL-ARCNN model. Using residual learning, the peak signal-to-noise ratio (PSNR) was the highest among all different image patch sizes with the best result of 38.09 dB in 50 by 50 patch size. 
	
	\begin{figure}
		\centering
		\noindent \includegraphics*[width=6.50in, height=4.20in, keepaspectratio=true]{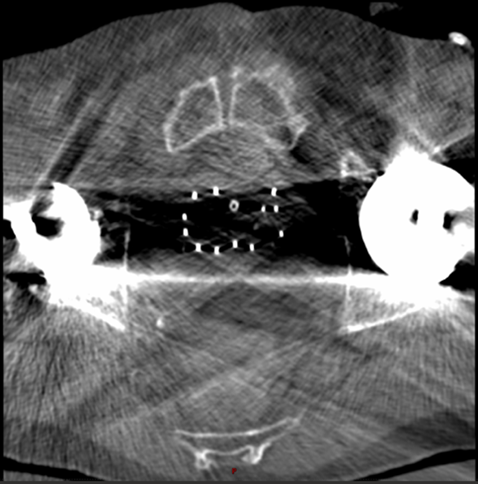}
		
		\noindent Figure 6. Example of metal artifacts caused by hip prothesis in a CT scan for prostate brachytherapy.
	\end{figure}
	
	Photoacoustic imaging, used for detecting prostate low-dose-rate (LDR) brachytherapy seeds, is susceptible to acoustic wave reflection artifacts . Allman et al. \cite{RN23} built a CNN to identify artifacts and true source (LDR seeds), and reduce the reflection artifact created by photoacoustic waves to improve the quality of seed reconstruction. The CNN model achieved a low localization error of point source with mean ± SD of 0.40 ± 0.22 mm and 0.38 ± 0.25 mm, and a high classification accuracy of 100
	
	Deep learning can also improve image quality by increasing resolution. In practice, 3D TRUS images with low resolution are usually captured with thick slice of (2-5 mm) in brachytherapy due to time constraint. He et al. \cite{RN53} developed a GAN-based framework integrated with a deeply supervised attention model to construct high-resolution ultrasound images from the low-resolution TRUS images. The model was trained with high and low-resolution TRUS image pairs from 20 patients, with the high-resolution images served as ground truths. The proposed model achieved a mean absolute error (MAE) of 6.5 ± 0.5 and a high PSNR of 38.0 ± 2.4 dB. 
	
	In addition to enhancing image resolution, deep learning models can also synthesize images, such as generating synthetic MRI from CT images, combing the cost efficiency of CT with the soft tissue contrast of MRI. Podgorsak et al. \cite{RN54} and Kang et al. \cite{RN133} both utilized Pix2Pix \cite{RN55} and CycleGAN to build DL-based models to create synthetic MRI from CT scans, namely PCGAN and PxCGAN. Both studies evaluated the quality of the synthetic MRIs by comparing contour differences between real and synthetic MRIs within the same radiation oncologist (inter-modality) and segmentation differences among different radiation oncologists within the same image (inter-observer). As presented in Table 4, both studies successfully visualized the prostate and catheters on the synthetic MRIs and showed that the DSC and MSD from inter-modality contours are comparable or better than the inter-observer differences. However, the limited patient dataset used in these studies may impact the generalizability of the models, which affect the image quality and corresponding dosimetric parameters on new synthetic MRIs. For instance, the dosimetric parameters for PTV and OARs from the synthetic MRI were generally higher than those from the actual MRI but had no statistically significant difference except for bladder $D_{1cc}$, potentially caused by the effect of catheter position on the target. Overall, this approach has the potential to improve clinical workflows by reducing the need for additional MRI scans.
	
	\begin{table}[]
		\centering
		\caption{AI implementation on other imaging applications}
		\label{tab:my-table}
		\resizebox{\textwidth}{!}{%
			\begin{tabular}{|p{3cm}|p{2cm}|p{2.5cm}|p{2cm}|p{2cm}|p{7cm}|p{2cm}|}
				\hline
				Application                & Cancer Site & Image Modality & Number of Patients & Model               & Result Summary                                                                                                                                                                                                                      & Citation \\ \hline
				Reduce metal artifacts     & GYN         & CT             & 35                 & RL-ARCNN            & PSNR = 33.83 dB, 38.09 dB, and 36.80 dB for 25 by 25, 50 by 50, and 100 by 100 image patch sizes respectively.                                                                                & \cite{RN21} \\ \hline
				Remove artifacts           & Prostate    & Photoacoustic   & 17,340 images      & CNN                 & From the water bath and phantom test, the mean localization error of point source were 0.40 ± 0.22 mm and 0.38 ± 0.25 mm.                                                                     & \cite{RN23} \\ \hline
				Improve resolution         & Prostate    & TRUS           & 20                 & GAN and Attention   & Mean MAE for reconstructed images = 6.5 ± 0.5 and PSNR = 38.0 ± 2.4 dB.                                                                                                                       & \cite{RN53} \\ \hline
				Synthetic MRI from CT      & Prostate    & CT - MRI       & 78                 & PCGAN               & \begin{tabular}[t]{@{}p{6cm}@{}} DSC = 0.852 ± 0.057 and MSD = 2.47 ± 0.50 mm for interobserver contour differences; \\ DSC = 0.846 ± 0.045 and MSD = 2.19 ± 0.69 mm for intermodality contour differences. \end{tabular}              & \cite{RN54} \\ \hline
				Synthetic MRI from CT      & Prostate    & CT - MRI       & 78                 & PxCGAN              & \begin{tabular}[t]{@{}p{6cm}@{}} DSC = 0.84 ± 0.05, MSD = 2.30 ± 0.67 mm, and HD = 10.11 ± 2.71 mm for interobserver contour differences; \\ DSC = 0.84 ± 0.05, MSD = 2.19 ± 0.69 mm, and HD = 8.34 ± 2.27 mm for intermodality contour differences. \\ MAE = 0.14 ± 0.03, MSE = 0.04 ± 0.01, PSNR = 68.69 ± 1.44 dB, and structural similarity index (SSIM) = 0.73 ± 0.11 for differences between synthetic MRI and real MRI. \end{tabular} & \cite{RN133} \\ \hline
				Synthetic MRI from CT      & Prostate    & CT - MRI       & 140                & CycleGAN and deep attention U-Net & \begin{tabular}[t]{@{}p{6cm}@{}} For synthetic MRI generated segmentations: DSC = 0.95 ± 0.03 and MSD = 0.52 ± 0.22 mm for bladder, \\ DSC = 0.87 ± 0.04 and MSD = 0.93 ± 0.51 mm for the prostate, \\ DSC = 0.89 ± 0.04 and MSD = 0.92 ± 1.03 mm for the rectum. \end{tabular} & \cite{RN149} \\ \hline
				Synthetic MRI from CT      & Prostate    & CT - MRI       & 49                 & CycleGAN            & \begin{tabular}[t]{@{}p{6cm}@{}} For synthetic MRI generated segmentations: DSC = 0.92 ± 0.09, HD = 4.38 ± 4.66 mm, and MSD = 0.62 ± 0.89 mm for the leave-one-out test. \\ DSC = 0.91 ± 0.07, HD = 4.57 ± 3.03 mm, and MSD = 0.62 ± 0.65 mm for the hold-out test. \end{tabular} & \cite{RN150} \\ \hline
			\end{tabular}%
		}
	\end{table}

	\noindent 
	\section{Preplanning}
	
	The preplanning process involves developing a brachytherapy treatment plan outside the operating room prior to its delivery to the patient. As summarized in Table 5, this process can be applied to both GYN and prostate brachytherapy treatments.

	\noindent 
	\subsection{GYN}
	For HDR GYN brachytherapy, preplanning involves selecting the appropriate applicators, either intracavitary (IC) or interstitial (IS), based on patient anatomy. Stenhouse et al. \cite{RN56} trained 12 machine learning models to predict the use of IC or IS applicator for different GYN cancer patients, selecting the top three models for final predictions. Important features, such as the needle patterns and clinical contours, were selected by an F-test to reduce complexity introduced by unnecessary features. An AdaBoost Classifier, a Gradient Boosting Classifier, and a Random Forest (RF) Classifier were selected to form a voting model because of their superior performances. The final voting model, formulated from the weighted average of the predicted probabilities from the three models, provides applicator decisions with an accuracy of 91.5 ± 0.9\% and a F1 Score of 90.6 ± 1.1\%. Another important factor to consider when selecting the appropriate applicator is soft-tissue deformation caused by applicator insertion. Applicators used in GYN brachytherapy can induce an average point-to-point displacement of 25.0 mm in the uterus. \cite{RN57} Knowing the changes in soft tissue location post-insertion gives physicians additional information for selecting the most effective applicator. Ghosh et al. \cite{RN57} used a deep-learning model to automatically predict uterus deformation caused by applicators using only pre-surgery MRI as input. The model utilized a deep CNN model with auto-encoders for pre-surgery uterus segmentation and a modified U-Net for predicting the uterus deformation during implantation. The predicted new uterine position had a DSC of 0.881 ± 0.038 and an HD of 5.8 ± 3.6 mm compared to the true deformed position. 
	
	\noindent 
	\subsection{Prostate}
	In LDR prostate brachytherapy, preplanning involves determining the seed distribution and the resulting dose distribution prior to the treatment day. Manual planning is time-consuming and heavily dependent on the experience level of the planner. Nicolae et al. [85] built a machine learning-based prostate implant planning algorithm (PIPA) system to automate treatment planning for LDR prostate patients. The percentages of plans that need minor or major modification were approximately the same for PIPA and manual method, but the algorithm reduced the planning time to 2.38 ± 0.96 minutes compared to 43.13 ± 58.70 minutes. However, this model was only tested on a small cohort of 41 patients and did not report differences in dose-volume histogram (DVH). In 2021, Aleef et al. \cite{RN63} employed a two-stage framework that consisted of conditional generative adversarial networks (cGAN) to automatically generate treatment plans. The cGAN model predicted needle distribution, and a simulated annealing algorithm optimized seed location. Overall, 90\% of the generated treatment plans were acceptable, with 60\% requiring minor modifications and 30\% requiring major modifications. Additionally, significantly less time (3 minutes compared to 20 minutes) was required for an automatic plan with similar quality to manual plans (a CTV $V_100$ value of 98.98\% compared to 99.36\%). In a later study \cite{RN64}, the researchers developed a treatment planning generative adversarial network (TP-GAN) and achieved similar DVH with less urethral doses in only 2.5 minutes or 3 seconds per plan with and without fine-tuning. Deep learning models are also potential substitutes for traditional computational models such as Monte Carlo (MC) simulations due to their fast-processing speed. Berumen et al. used a 3D U-Net based model that learns MC simulations to predict single-seed dose to medium ($D_{M,M}$) on CT images, which produced similar DVH metrics but in significantly less time than MC methods (1.8 ms compared to 2 hours). 
	
	In intraoperative LDR brachytherapy, extra seeds are usually ordered to ensure sufficient coverage. However, unused seeds in the procedure would require physicist to spend additional time documenting and safely returning them to the vendor. To address this, Boussion et al. \cite{RN60} used several machine-learning models to predict the number of seeds needed for LDR prostate brachytherapy. The best-performing ML model, support vector machines for regression (SVR), reduced the unused seeds from 23 ± 4 to 10 ± 4 when tested on 38 unseen treatments, though a 10\% margin is still necessary to prevent seed depletion during implant. 
	
	\begin{table}[]
		\caption{AI application in preplanning}
		\label{tab:my-table}
		\resizebox{\textwidth}{!}{%
			\begin{tabular}{|p{3cm}|p{2cm}|p{3cm}|p{2cm}|p{7cm}|p{2cm}|}
				\hline
				Preplanning Task & Cancer Site & Number of Patients & Model & Result Summary & Citation   \\ \hline
				Select Applicator & GYN & 233 & AdaBoost, Gradient Boost, and RF classifier & Accuracy = 91.5 ± 0.9\% and F1 Score = 90.6 ± 1.1\%. & \cite{RN56}  \\ \hline
				Predict Applicator Induced Uterine Deformation & GYN & 92 & CNN and modified U-Net & For predicted uterine position after applicator insertion: DSC = 0.881 ± 0.038 and HD = 5.8 ± 3.6 mm. & \cite{RN57}  \\ \hline
				Generate Preplans & Prostate & 150 & K-nearest neighbor & No significant difference in prostate $V_{100}$, prostate $D_{90}$, urethra $D_{0.1cc}$, rectum $D_{1cc}$, PTV $V_{100\%}$ between manual and ML plans, except prostate $V_{150\%}$ was 4\% lower for ML plans. & \cite{RN66}   \\ \hline
				Generate Preplans & Prostate & 41 & PIPA & No significant differences in prostate $D_{90\%}$, $V_{100\%}$, rectum $V_{100\%}$, or rectum $D_{1cc}$ between manual and PIPA plans. & \cite{RN67}  \\ \hline
				Predict Needles and Seeds Distribution & Prostate & 931 & cGAN & 98.98\% achieved 100\% of the prescribed dose, 90\% of the generated plans were acceptable with 60\% minor and 30\% major modifications. PTV $V_{100\%}$ = 96.55 ± 1.44, PTV $V_{150\%}$ = 56.23 ± 4.37, CTV $V_{100\%}$ = 99.36 ± 0.96, CTV $V_{150\%}$ = 63.03 ± 5.15. & \cite{RN63}   \\ \hline
				Predict Seeds Distribution & Prostate & 961 & TP-GAN & 98.98\% achieved 100\% of the prescribed dose. TPGAN only: PTV $V_{100\%}$ = 94.6 ± 3.9, PTV $V_{150\%}$ = 55.0 ± 11.9, CTV $V_{100\%}$ = 97.8 ± 2.5, CTV $V_{150\%}$ = 60.8 ± 13.7. TPGAN and Simulating Annealing: PTV $V_{100\%}$= 95.9 ± 1.6, PTV $V_{150\%}$ = 53.0 ± 3.5, CTV $V_{100\%}$= 98.8 ± 0.9, CTV $V_{150\%}$ = 59.1 ± 5.0. & \cite{RN64} \\ \hline
				Predict Single-Seed Dose to Medium in Medium & Prostate & 44 & 3D U-Net CNN & The average differences of the predicted and MC-based calculations were 0.1\% for CTV $D_{90}$ and 1.3\%, 0.07\%, and 4.9\% for the $D_{2cc}$ of rectum, bladder, and the urethra. & \cite{RN65}   \\ \hline
				Predict Number of Seeds & Prostate & 409 & SVR & MSE = 2.55, MAE = 1.21, and maximum error = 7.29 seeds. When tested on 38 unseen treatments, reduced unused seeds from 23 ± 4 to 10 ± 4 seeds, and saved 493 seeds in total. & \cite{RN60}   \\ \hline
			\end{tabular}%
		}
	Note: Abbreviations: MSE (mean squared error). 
	\end{table}
	
	\noindent 
	\section{Treatment Planning}
	In HDR brachytherapy, both prostate and GYN, the treatment planning procedure is often performed intraoperatively with patients under anesthesia or immobilized, which makes it highly time sensitive. The detailed methods and results of recent studies employing AI to support the treatment planning process are summarized in Table 6.
	
	\noindent 
	\subsection{Prostate}
	
	In HDR brachytherapy, the dose distribution largely depends on the positions of applicators, which vary per patient, making the dose prediction challenging. Figure 7 is an example of prostate HDR treatment planning performed intraoperatively on TRUS. AI can analyze large datasets and identify complex patterns, allowing it to predict DVH for the CTV and OARs. \cite{RN71, RN73, RN70, RN74, RN68, RN75, RN77, RN136, RN69, RN72}
	
	\begin{figure}
		\centering
		\noindent \includegraphics*[width=6.50in, height=4.20in, keepaspectratio=true]{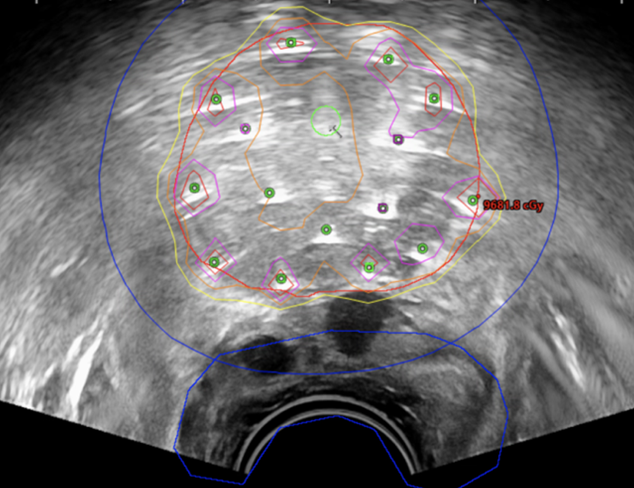}
		
		\noindent Figure 7. Example of HDR prostate brachytherapy treatment planning on TRUS. 
	\end{figure}

	Mao et al. \cite{RN75} trained a 3D deep convolutional neural network (DCNN), RapidBrachyDL, to perform dose calculations based on the MC method. RapidBrachyDL had errors less than 1.5\% for the CTV and OAR DVH metrics as shown in Table 6, comparable to the conventional MC method. Interestingly, although trained with prostate patient data only, it showed transferability to cervical cancer patient CT scans with errors below 3\%. Similarly, Villa et al. \cite{RN77} trained a DCNN model with MC-based method generated database, achieving fast inverse planning in only 1.7 seconds with small mean percent errors (MPE). 
	
	To guide the placement of catheters in HDR prostate brachytherapy, Lei et al. \cite{RN59} developed Reg-Net, a 3D CNN based deformable multi-atlas registration network. Reg-Net used distance maps of targets and OARs to register with new patient CT images and predict catheter locations. The resulting DVH metrics from predicted catheter distributions differed by no more than 5\% different from clinical plans, though there were hot spots in the prostate and excessive dose in OARs. This model quantifies dose distribution on CT simulations prior to catheter placements, locates catheters more effectively, which potentially reduces intuitive decision-making in the HDR procedures. 
	
	\noindent 
	\subsection{GYN}
	For HDR GYN brachytherapy, knowledge-based planning (KBP) using traditional methods has been proven to standardize treatment plans and effectively predict dose volumes by learning the correlation between the final plan dosimetry and patient-specific characteristics \cite{RN76}. KBP uses spatial information of patient anatomy and applicators to predict 3D dose for high-quality treatment plans. Cortes et al. \cite{RN70} applied a 3D U-Net CNN to conduct KBP for HDR cervical cancer brachytherapy using tandem and ovoids (T\&O) applicators. This model provided objective dose measures for HR-CTV $D_{90}$ and OAR $D_{2cc}$, potentially offering quality checks for planners. MC-based dose calculation can also be performed with DL models. Akhavanallaf et al. \cite{RN71} developed a deep neural network (DNN)-based model, personalized brachytherapy dosimetry simulator (PBrDoseSim), to calculate dose with the MC method. PBrDoseSim predicted the specific energy deposition kernel when the radioactive source is positioned at the dwell position, which exhibited good agreement with MC calculations. Additionally, the study provided a baseline comparison and an evaluation of their full dose distribution against the ground truth as shown in Table 6. The study is limited by using the contours instead of the direct density maps from the CT images. 
	
	To increase model robustness against different GYN brachytherapy settings, Li et al. \cite{RN68} trained a Squeeze and Excitation Attention Net ($SE_{AN}$) with various clinically common applicators including vaginal applicator, ovoid applicator, T\&O applicators, free needles, and multi-channel applicator. The smallest MAE of the predicted DVH metrics from $SE_{AN}$, U-Net, and 3D Cascaded U-Net were summarized in Table 6. While $SE_{AN}$ did not outperform U-Net and Cascaded U-Net in rectum and HRCTV MAEs, it predicted the correct number and dose gradient for each applicator setup and closely resembled the ground truth and had the highest gamma passing rate of 92 ± 10
	
	Several studies showed AI model’s capability in predicting OAR dose \cite{RN73, RN74, RN69, RN72}. Two studies \cite{RN73, 72} predicted the rectal toxicity levels from treatment plans, where the best sensitivity, or recall, is 84.75\%, and the best specificity, or precision, is 79.87\%. When using statistically significant features only, the model achieved an area under the receiver operating characteristic curve (AUC) of 0.91, indicating its strong discriminative ability. Zhang et al. \cite{RN69} predicted the dose in bladder, rectum, and sigmoid using neural networks, correlating $D_{2cm^3}/D_{90}$ for each OAR and its sub-organ, showing strong correlation and small mean squared error (MSE). Additionally, this model only requires sub-organ volumes of the OAR without additional voxel information, allowing physicians without programming experience to perform dose predictions.
	
	Besides dose prediction, AI has several other applications in enhancing the treatment planning of HDR brachytherapy. For example, Gao et al. \cite{RN119} used dose prediction network (DPN) and a plan-approval probability network (PPN) to predict the HDR plan approval probability. Pu et al. \cite{RN109} provided a method, the intelligent treatment planner network (ITPN) to automatically adjust HDR source dwell times, optimizing until the objective function converges. Compared to a common clinical model, inverse planning simulated annealing (IPSA), the ITPN model preserved OARs better, notably reducing the bladder $D_{2cc}$,  rectum $V_{150}$, sigmoid $V_{150}$, rectum $V_{200}$, and sigmoid $V_{200}$.  Shen et al. \cite{RN110} used deep reinforced learning-based optimization to provide automatic weight tuning in inverse treatment planning. The method consisted of a Weight Tuning Policy Network (WTPN) which automatically adjust the weights of OARs to produce high-quality plans, showing proficiency even trained on a limited dataset and potential for integration into other treatment planning optimization.
	
	{
		\small
		\begin{longtable}{|p{2cm}|p{1.5cm}|p{2cm}|p{2cm}|p{6cm}|p{2cm}|}
			\caption{AI application in treatment planning}\\
					\hline
					Treatment Planning Task & Cancer Site & Number of Patients & Model & Results & Citation \\ \hline
					Inverse Planning & Prostate & 273 & DCNN & For prostate: MPE = -0.8 ± 1.0 \% for $V_{100}$, -0.6 ± 3.1\% for $V_{150}$, and 0.2 ± 1.3\% for $V_{200}$. For OARs: MPE = 1.7 ± 3.5\% for urethra $D_{10}$, 0.9 ± 2.7\% for urethra $D_{30}$; 0.4 ± 2.6\% for rectum $D_{2cc}$, 2.8 ± 9.2\% for rectum $D_{0.1cc}$. & \cite{RN77} \\ \hline
					Predict Catheter Placements & Prostate & 90 & Reg-Net & The difference between the clinical and predicted prostate $V_{150}$, $V_{200}$, and $D_{90}$ were 5.0 ± 6.5\%, 2.9 ± 4.3\%, 0.9 ± 1.5\%, bladder $D_{2cc}$, $V_{75}$ were 3.5 ± 3.4\% and 0.2 ± 0.4 cc, rectum $D_{2cc}$, $V_{75}$ were 1.5 ± 4.1\% and 0.1 ± 0.4 cc, urethra $V_{125}$= 0.3 ± 0.4cc. & \cite{RN59} \\ \hline
					CTV and OAR dose prediction & Prostate/GYN & 61 & RapidBrachyDL & For prostate cancer: dose prediction errors = 0.73\%, 1.1\%, 1.45\%, 1.05\%, for CTV  $D_{90}$, rectum $D_{2cc}$, urethra $D_{0.1cc}$, and bladder $D_{2cc}$; For cervical cancer: dose prediction errors = 1.73\%, 2.46\%, 1.68\%, and 1.74\% for CTV $D_{90}$, rectum $D_{2cc}$, sigmoid $D_{2cc}$, and bladder $D_{2cc}$, respectively. & \cite{RN75} \\ \hline
					CTV and OAR dose prediction & GYN & 126 & 3D U-Net CNN & Isodose DSC = [0.87, 0.94], mean difference of the DVH metrics were -0.09 ± 0.67 Gy for HRCTV $D_{90}$, -0.17 ± 0.67 Gy for bladder $D_{2cc}$, -0.04 ± 0.46 Gy for rectum $D_{2cc}$, and 0.00 ± 0.44 Gy for sigmoid $D_{2cc}$. & \cite{RN70} \\ \hline
					CTV and OAR dose prediction & GYN & 78 & PBrDoseSim & For model predicted single-dwell dose kernels, MRAE = 1.16±0.42\% MAE=4.2±2.7x$10^{-4}$ ($Gy.sec^{-1}$/voxel). Conformity index = 0.24, dose non-uniformity ratio = 0.65, and dose homogeneity index = 0.34. The MRAE for CTV between DNN and MC were 1.5±0.88\% for $D_{95}$, 1.8±0.86\% for $D_{90}$, 1.3 ± 1\% for $D_{50}$, 0.85 ± 0.43\% for $V_{200}$, 0.56 ± 0.56\% for $V_{150}$, 1.48 ± 0.72\% for $V_{100}$, 0.26 ± 0.38\% for $V_{50}$. The MRAE for OARs were 2.7 ± 1.7\% and for bladder $D_{5cc}$ and $D_{2cc}$, 1.9 ± 1.3\% and 2.4 ± 1.6\% for sigmoid $D_{5cc}$ and $D_{2cc}$, and 2.1 ± 1.7\% and 2.5 ± 2\% for rectum $D_{5cc}$ and $D_{2cc}$. & \cite{RN71} \\ \hline
					CTV and OAR dose prediction & GYN & 81 & $SE_AN$ & $SE_AN$: MAE = 0.37 ± 0.25 for HRCTV $D_{90}$, 0.23 ± 0.14 for bladder $D_{2cc}$, 0.28 ± 0.20 for rectum $D_{2cc}$. U-Net: MAE = 0.34 ± 0.24 for HRCTV $D_{90}$, 0.25 ± 0.20 for bladder $D_{2cc}$, 0.25 ± 0.21 for rectum $D_{2cc}$. Cascaded U-Net: MAE = 0.42 ± 0.31 for HRCTV $D_{90}$, 0.24 ± 0.19 for bladder $D_{2cc}$, 0.23 ± 0.19 for rectum $D_{2cc}$. & \cite{RN68} \\ \hline
					CTV and OAR dose prediction & GYN & 224 & 3D mask-guided dose prediction model & Dose prediction errors = 0.63 ± 0.63, 0.60 ± 0.61, 0.53 ± 0.61, 1.21 ± 0.85, 0.71 ± 0.61, 1.16 ± 1.09, and 0.86 ± 0.58, for HRCTV $D_{95}$, HRCTV $D_{95}$, HRCTV $D_{100}$, bladder $D_{2cc}$, sigmoid $D_{2cc}$, rectum $D_{2cc}$, and intestine $D_{2cc}$. & \cite{RN136} \\ \hline
					Rectum dose prediction & GYN & 42 & VGG-16 and RSDM & 10-fold cross validation: sensitivity = 61.1\%, specificity = 70\%, and AUC = 0.7. leave-one-out cross validation: sensitivity = 75\% specificity = 83.3\%, and AUC = 0.89. & \cite{RN72} \\ \hline
					Rectum dose prediction & GYN & 42 & SVM & Using principal component analysis (PCA) features: sensitivity = 74.75\%, specificity = 72.67\%, and AUC = 0.82; Using statistically significant features: sensitivity = 84.75\%, specificity = 79.87\%, and AUC = 0.91. & \cite{RN73} \\ \hline
					OAR dose prediction & GYN & 59 & LM algorithm & R= 0.80 for bladder, 0.88 for rectum, and 0.86 for sigmoid; MSE = 5.543×$10^{-3}$ for bladder $D_{2cc}/D_{90}$, 5.028x$10^{-3}$ for rectum $D_{2cc}/D_{90}$, and 8.815x$10^{-3}$ for sigmoid $D_{2cc}/D_{90}$. & \cite{RN69} \\ \hline
					Predict Plan Approval Probability & GYN & 63 & DPN and PPN & Dose prediction error = 11.51\% ± 6.92\% for bladder, 8.23\% ± 5.75\% for rectum, 7.12\% ± 6.00\% for sigmoid $D_{2cc}$, and 10.16\% ± 10.42\% for CTV $D_{90}$. Plan approval prediction: accuracy = 0.70, sensitivity = 0.74, specificity = 0.65, and AUC = 0.74. & \cite{RN119} \\ \hline
					Inverse planning - dwell times & GYN & 20 & ITPN & Directly output the dwell times of preselected dwell positions of HDR BT for cervical cancer, generate higher quality plans with better CTV coverage and OARs sparing compared to clinically accepted IPSA. & \cite{RN109} \\ \hline
					Organ Weighting Factor Adjustment & GYN & 10 & WTPN & Plan quality score was improved by 8.5\% compared to the initial plan with arbitrarily set weights, and by 10.7\% compared to the plans generated by human planners. & \cite{RN110} \\ \hline
					Intra-fractional OARs dose-volume histogram prediction & GYN & 30 & ANN & MPE = 6\%, 5\%, 8\%, 7\%, 10\% for predicting intra-fractional dose variations of bladder, rectum, sigmoid, $CTV_{IR}$, and $CTV_{HR}$, respectively. & \cite{RN74} \\ \hline
					Optimizing Treatment Parameters & Prostate & 35 & MANN and GA & The optimal values for Therapy Dose (TD) = 47.3 Gy, TD coverage index ($CI_{100\%}$) = 1.14, and PSA nadir = 0.047 ng $cm^{-3}$ for low-risk group; TD = 50.4 Gy and $CI_{100\%}$ = 1.6, and PSA nadir = 0.25 ng $cm^{-3}$ for high-risk group. & \cite{RN153} \\ \hline
			
		\end{longtable}
	Note: Abbreviations: MRAE (mean relative absolute error), VGG-16 (16-layers CNN developed by the visual geometry group), SVM (support vector machine), RSDM (rectum surface dose maps), LM (Levenberg–Marquardt), ANN (artificial neural network), MANN (multilayer artificial neural network), GA (genetic algorithm), MO-RV-GOMEA (multi-objective real-valued gene-pool optimal mixing evolutionary algorithm), BRIGHT (brachytherapy via artificial Intelligent GOMEA-Heuristic based treatment planning), Rad-TRaP (radiomics based targeted radiotherapy planning).
	}

	\bigbreak
	
	\noindent 
	\section{Applicator Reconstruction}
	
	\noindent 
	\subsection{GYN}
	
	Accurate digitization of applicators and catheters is crucial in HDR GYN treatment planning as it largely affect the final dose distribution. Table 7 provides a summary of the AI implementation in different types of applicator reconstruction, such as the T\&O, tandem \& ring (T\&R), fletcher applicators, catheters, and needles. To automatically segment the T\&O applicators on CT images, Jung et al. \cite{RN33} trained a U-net model with additional spectral clustering and polynomial curve-fitting methods to identify the locations and the central paths of applicators. Although trained solely on CT images T\&O applicators, the model accurately digitized not only T\&O applicators but also Y-tandem and cylinder applicators on CT, as well as T\&O applicators on cone-beam CT.  Both tip position errors and HD were less than 1 mm under various testing setup, indicating the model's robustness and transferability when applied to different applicator structures and lower image quality.
	
	MRI provides better visibility of soft tissue but has difficulty imaging applicators due to the inconsistent appearance of contrast fiducials on the applicator lumen. However, HDR cervical brachytherapy applicators can be reconstructed using MRI alone with library applicator models from the manufacturers, which is potentially more reproducible than digitization on CT images. Hrinvich et al. \cite{RN38} utilized a circular Hough transform (CHT) model \cite{RN37} to identify tandem and ring applicators on T2-weighted MR images. It is followed by a 3D rotation matrix and a 3D translation vector with a stochastic evolutionary optimizer \cite{RN36} to obtain the positions of T\&R applicators with a rigid registration method. The proposed method achieved a mean reconstruction accuracy of 0.60 ± 0.24 mm for the ring and 0.58 ± 0.24 mm for the tandem, with mean variability smaller than the inter-observer variability. Plastic catheters have a low hydrogen composition, which makes them difficult to identify on the MRI. Zaffino et al. \cite{RN35} developed a 3D U-Net model to automatically segment multiple closely spaced brachytherapy catheters in MRI. The model produced a good accuracy with an average distance error of 2.0 ± 3.4 mm, DSC of 0.60 ± 0.17, and a HD of 15.9 ± 20.5 mm. The false positive and false negative catheters were 6.7\% and 1.5\%, respectively. Similar results were provided by Shaaer et al. \cite{RN161}, as outlined in Table 7. The accuracy of the segmentation struggled with tubular structure reconstruction, especially on MR images where other tubular tissue structures exist.
 
	\begin{table}[]
		\caption{AI in GYN applicator reconstruction }
		\label{tab:my-table}
		\resizebox{\textwidth}{!}{%
			\begin{tabular}{|p{2cm}|p{1.5cm}|p{2cm}|p{2cm}|p{6cm}|p{2cm}|}
				\hline
				Type of Applicator & Image Modality & Number of Patients & Model & Results & Citation \\ \hline
				T\&O applicator & CT & 13 & U-Net & For applicator segmentation: DSC = 0.937 ± 0.064 in 2D and 0.931 in 3D. HD = 1.01 mm, 1.21 mm, and 1.18 mm. For tip position errors: tandem = 0.5 mm, right ovoid 0.74 mm, and left ovoid = 0.67 mm. & \cite{RN33} \\ \hline
				T\&O applicator & CT & 91 & DSD-UNET & For the intrauterine and ovoid tubes: DSC = 0.921, HD = 2.3 mm. Between channel paths: HD = 0.88 ± 0.12 mm, 0.95 ± 0.16 mm, and 0.96 ± 0.15 mm for the intrauterine, left ovoid, and right ovoid tubes, respectively. & \cite{RN41} \\ \hline
				T\&O applicator & CT & 10 & HDBSCAN & Mean contour error = 0.3 mm, HD<=1 mm. For HR-CTV $D_{90}$, HR-CTV $D_{95}$, bladder $D_{2cc}$, rectum $D_{2cc}$, large bowel $D_{2cc}$, and small bowel $D_{2cc}$, the median and mean difference in DVH parameters are all <= 1\%. & \cite{RN156} \\ \hline
				Fletcher applicator & CT & 70 & U-Net & For applicator segmentation: DSC = 0.89 ± 0.09, HD = 1.66 ± 0.42 mm, shaft error < 0.5 mm, tip error = 0.8 mm. Dosimetric differences = 0.29\% for the $D_{90}$ of HRCTV, and less than 2.64\% for OAR $D_{2cc}$. & \cite{RN34} \\ \hline
				T\&R applicator & MRI & 10 & CHT & For dwell positions, the mean errors = 0.60 ± 0.24 mm and 0.58 ± 0.24 mm, variability = 0.72 ± 0.32 mm and 0.70 ± 0.29 mm, and inter-observer variability = 0.83 ± 0.31 mm and 0.78 ± 0.29 mm, for the ring and tandem, respectively. & \cite{RN38} \\ \hline
				Catheters & MRI & 50 & 3D U-Net & DSC = 0.60 ± 0.17, HD = 15.9 ± 20.5 mm, MDE = 2.0 ± 3.4 mm, false positive = 6.7\%, and false negative = 1.5\%. & \cite{RN35} \\ \hline
				Catheters & MRI & 20 & 2D U-Net & DSC = 0.59 ± 0.10 and HD = 4.20 ± 2.40 mm. Average variation = 0.97 ± 0.66 mm with 98.32\% < 2 mm and 1.68\% >= 3 mm & \cite{RN161} \\ \hline
				Interstitial needles & CT & 15 & 2.5D U-Net & DSC = 0.93 for needle segmentation, HD = 0.71 mm for needle trajectories, and HD = 0.63 mm for needle tip positions. & \cite{RN157} \\ \hline
				Interstitial needles & CT & 17 & 3D Unet with attention gates & DSC = 0.937 ± 0.014, JI = 0.882 ± 0.025, HD =3.0 ± 1.9 mm, tip error = 1.1 ± 0.7 mm, and shaft error = 1.8 ± 1.6 mm & \cite{RN158} \\ \hline
				Needle structures & Ultrasound & 5 prostate patients, 6 GYN patients & CNN with modified U-Net & For prostate: needle tip error = 1.5 [0.9, 8.3] mm, angular error = 0.4 [0.3, 0.7] °, HD = 6.8 [1.3, 17.6] mm, DSC = 0.789 [0.738, 0.847], recall = 73.2 [62.4, 81.9] \%, and precision = 87.9 [84.8, 95.8] \%. For GYN: needle tip error = 0.3 [0.2, 0.4] mm, angular error = 0.4 [0.2, 0.7] °, HD = 0.5 [0.4, 0.9] mm, DSC = 88.7 [84.6, 93.5] \%, recall = 85.2 [80.9, 91.1] \%, and precision = 93.2 [89.6, 97.0] \%. & \cite{RN159} \\ \hline
				T\&R applicator and interstitial needles & CT & 48 & nnU-Net and 3D U-Net & DSC = 0.646 for T\&R applicators and 0.738 ± 0.034 for interstitial needles. & \cite{RN160} \\ \hline
			\end{tabular}%
		}
	Note: Abbreviations: HDBSCAN (density-based linkage clustering algorithm), DSD-UNET (dilated convolution and deep supervision U-Net). Duplicate paper from Zhang et al. \cite{RN41} as in Table 2.
	\end{table}
	
	\noindent 
	\subsection{Prostate}
	For prostate brachytherapy, both HDR and LDR, are unique procedures with comparable workflow and distinct advantages \cite{RN121}. As detailed in Table 8, approximately the same number of studies used deep learning to assist HDR applicator reconstruction (identifying needles and catheters) and LDR (reconstructing needles and seeds under different imaging modalities). HDR brachytherapy for prostate cancer involves the insertion of interstitial needles through the perineum, followed by imaging using techniques such as TRUS, MRI, or CT. TRUS is the most common imaging modality used to guide the insertion of needles for HDR prostate brachytherapy. Andersén et al. \cite{RN25} used a 3D CNN U-Net model to identify needles in TRUS images. The model was trained on a large dataset consisting of 1102 brachytherapy treatments, with a total of 24422 individual needles. The model achieved a root-mean-square distance (RMSD) of 0.55mm compared to the clinical ground truth and 0.75 mm compared to another physicist’s digitization, which is lower than the inter-observer variability of 0.80 mm. To further enhance the needle digitization workflow, Zhang et al. conducted three studies using different deep-learning approaches to detect multiple needles on 3D TRUS images simultaneously. The three approaches were using an order-graph regularized dictionary learning (ORDL)-based method \cite{RN26}, a deeply supervised attention U-Net with a total variation (TV) regularization method \cite{RN27}, and a large margin mask R-CNN model (LMMask R-CNN) with a needle-based density-based spatial clustering method \cite{RN28}. The shaft and tip errors and accuracies were similar in all three methods but the LMMask R-CNN-based model had the most superior result, detecting 98\% of needles with shaft and tip errors of 0.091 ± 0.043 mm and 0.33 ± 0.363 mm, respectively. 
	
	\begin{figure}
		\centering
		\noindent \includegraphics*[width=6.50in, height=4.20in, keepaspectratio=true]{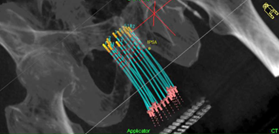}
		
		\noindent Figure 8. Needle reconstruction on CT image for HDR prostate brachytherapy.
	\end{figure}

	CT images can also be used to reconstruct the needles, as illustrated in Figure 8. Weishaupt et al. \cite{RN24} developed a deep-learning method to automatically digitize the HDR prostate needles on CT images. Using 2D U-Net architecture, the model segmented the catheters and reconstructed their geometries in 3D using a density-based linkage clustering algorithm. The model accurately digitized all needles in under one minute with a mean tip distance difference of -0.1 ± 0.6 mm and a mean shaft distance of 0.13 ± 0.09 mm. This method is highly efficient when compared to manual approach, which typically takes an average of two minutes per needle. 
	
	MRI-guided HDR prostate brachytherapy has the potential to optimize the dose distribution due to MRI's superior visualization of the DIL compared to TRUS or CT. \cite{RN171} However, digitizing catheters in MRI is challenging because of their dark and diffuse appearance. To address this issue, Dai et al. \cite{RN32} utilized an attention-gated U-Net to automatically digitize catheters and a TV regularization to remove excessive noise. The catheter tip error was found to be 0.37 ± 1.68 mm, with 87\% of the tips within localization error of no greater than 2.0 mm. Regarding catheter shaft localization, the error was 0.93 ± 0.50 mm, with 97\% of catheters detected with an error of less than 2.0 mm. The precision, recall, and F1 score of shaft localization were 0.96, 0.86, and 0.91, respectively. These results indicate that AI can simplify catheter digitization on MRI, potentially facilitate the use of MRI-guided brachytherapy in clinical practice and leading to improved patient outcomes.
	
	During the LDR prostate brachytherapy procedures, the placement of the radioactive seeds is guided by TRUS and often updated based on real-time changes in the prostate. It is crucial to compare the locations and orientations of implanted seeds to the planned seeds to adjust plans accordingly before the next seed implantation. Golshan et al. \cite{RN30} implemented a CNN model with a coarse sliding window approach to identify the needle tracks, then detect stranded seeds within each identified track. It achieved a precision of 78 ± 8\%, recall of 64 ± 10\%, and an F1 score of 70 ± 8\%. The relatively low recall suggests that many seeds were not identified on TRUS images, primarily due to poor ultrasound image quality and a small training cohort. Due to similar challenges, Holupka et al. \cite{RN102} was also not able to provide high identification precision for loose seeds on TRUS. 
	
	CT is commonly used to assess LDR treatment quality post-implant. Nguyen et al. \cite{RN31} used several machine learning-based models to segment single seeds and groups of closely spaced (union) seeds on CT images. They first used a k-means method to categorize the seeds into single or union seeds groups, then a modified k-means for seeds (k-means-FS) and a Gaussian mixture model (GMM) with expectation-maximization (EM) method to separate the union seed groups. The training and testing dataset comprised 14 patients with a total of 1063 implanted seeds, along with two phantoms (1 seed and 73 seeds). The seed angular orientation errors and MDE were greater in patients compared to the phantoms due to the complex anatomical structures, and the false detection rate ranged from 1.8\% to 4.8\% for different setups. 
	
	The implanted radioactive seeds can also be identified on MRI only, as it offers superior soft tissue contrast than CT. However, it can be challenging since they do not produce MR signals. An endorectal coil (ERC) can enhance seed visibility on MRI scans but may entail additional costs and discomfort for patients. To reduce the manual labor of identifying radioactive seeds on MRI after LDR brachytherapy, Sanders et al. \cite{RN29} developed a sliding-window CNN algorithm (SeedNet), which automatically identifies implanted radioactive seeds on prostate MRI scans. SeedNet demonstrated superior accuracy and reduced identification time compared to results from dosimetrists for patients both with and without an ERC. With ERC, seed detection achieved the highest recall (97.6\%), precision (97.2\%), and F1 score (97.4\%). Additionally, detection exhibited a low false discovery rate (2.8\%), low false-negative rate (2.4\%), and low root mean square error (RMSE) of 0.19 mm ± 0.04 mm, all within an inference time of 56.6 ± 18.2 seconds.
	
	\begin{table}[]
		\caption{AI in prostate applicator reconstruction}
		\label{tab:my-table}
		\resizebox{\textwidth}{!}{%
			\begin{tabular}{|p{2cm}|p{1.5cm}|p{2cm}|p{2cm}|p{7cm}|p{2cm}|}
				\hline
				Type of Applicator & Image Modality & Number of Patients & Model & Result Summary & Citation \\ \hline
				Needles & TRUS & 1102 & 3D CNN U-Net & RMSD = 0.55 [0.35 0.86] mm & \cite{RN25} \\ \hline
				Needles & TRUS & 21 & ORDL & Shaft error = 0.19 ± 0.13 mm, tip error = 1.01 ± 1.74 mm, accuracy = 0.95 & \cite{RN26} \\ \hline
				Needles & TRUS & 23 & Deep supervised attention U-Net & Shaft error = 0.29 ± 0.23 mm, tip error = 0.44 ± 0.93 mm, accuracy = 0.96 & \cite{RN27} \\ \hline
				Needles & TRUS & 23 & LMMask R-CNN & Shaft error = 0.091 ± 0.043 mm, tip error = 0.330 ± 0.363 mm, accuracy = 0.98 & \cite{RN28} \\ \hline
				Needles & TRUS & 823 & Modified deep U-Net and VGG-16 & Needle trajectories: resolutions = 0.668 mm and 0.319 mm in x and y direction. Needle tips: resolution = 0.721 mm, 0.369 mm, and 1.877 mm in x, y, and z directions. & \cite{RN162} \\ \hline
				Catheters & TRUS & 242 & U-Net and 3D reconstruction & 80\% of catheter reconstructions were accurate within 2 mm along 90\% of the catheter lengths. Reconstruction accuracy = 91\% excluding 27\% outliers. & \cite{RN163} \\ \hline
				Catheters & TRUS & 49 & 3D U-Net & AUC = 0.85, recall = 0.97, and true positive rate = 0.95. & \cite{RN164} \\ \hline
				Needles & CT & 57 & U-Net & Mean tip distance difference = -0.1 ± 0.6 mm with range from -1.8 mm to 1.4 mm. Mean shaft distance = 0.13 ± 0.09 mm with maximum distance = 0.96 mm. & \cite{RN24} \\ \hline
				Catheters & MRI & 20 & Deep supervised attention U-Net & Precision = 0.96, recall = 0.86, F1 = 0.91. Catheter tips localization: error = 0.37 ± 1.68 mm, 87\% with errors less than ± 2.0 mm, and more than 71\% within 1.0 mm. Catheter shaft localization: error = 0.93 ± 0.50 mm, 97\% with errors less than ± 2.0 mm, and 63\% within 1.0 mm. & \cite{RN32} \\ \hline
				Catheters & MRI & 35 & 3D U-Net & F1 score = 0.73, precision = 0.65, recall = 0.85, percentage of needles detected = 97.5\% & \cite{RN140} \\ \hline
				Stranded Seeds & TRUS & 13 & CNN & Precision = 78 ± 8\%, recall = 64 ± 10\%, F1 score = 70 ± 8\% & \cite{RN30} \\ \hline
				Needles & TRUS & 9 & Bayesian classifier & 23 fully visible needles: mean tip error = 1.4 mm, mean axis error = 1.5$^o$.19 partially visible needles: mean tip error = 4.2 mm, mean axis error = 6$^o$. & \cite{RN169} \\ \hline
				Stranded and Loose Seeds & TRUS & 1 phantom & Bayesian classifier and SVM & Mean error = 1.09 mm ± 0.61 mm on phantom image and 1.44 ± 0.45 mm on clinical images. Angle orientation error = 4.33 ± 8.5$^o$. & \cite{RN166} \\ \hline
				Loose Seeds & TRUS & 10 & DetectNet & Location error = 2.29 mm, precision = 81.07\% and recall = 82.29\%. & \cite{RN102} \\ \hline
				Loose Seeds & CT & 14 patients and 1 phantom & k-means, k-means-FS, and GMM with EM method & For the phantom, the angular orientation error = 0.96 ± 0.4°, and MDE = 0.08 ± 0.04 mm. For the patient, the maximum angular orientation error = 3.18 ± 0.9°, and MDE = 0.50 ± 0.16 mm. The least false detection rate = 1.8\% & \cite{RN31} \\ \hline
				Stranded Seeds with Positive Contrast MR-signal Seed Markers & MRI & 68 & SeedNet & With ERC: seed localization recall = 97.6 ± 2.2\%, precision = 97.2 ± 1.9\%, F1 score = 97.4 ± 1.5\%, and RMSE = 0.19 ± 0.04 mm. Without ERC: seed localization recall = 96.5 ± 2.3\%, precision = 90.5 ± 4.6\%, F1 score = 93.3 ± 3.0\%, and RMSE = 0.24 mm ± 0.03 mm. & \cite{RN29} \\ \hline
				Stranded Seeds & MRI & 1 phantom & QSM, k-means, and k-medoids clustering & The average length and width of detected seeds = 4.6 ± 0.3 mm and 0.9 ± 0.2 mm, compared with the actual dimensions of 4.5 mm and 0.8 mm. Maximum seed centroids difference = 7 mm. Dose distribution differences range from -8 to 8 Gy/U. & \cite{RN170} \\ \hline
			\end{tabular}%
		}
	Note: Abbreviations: VGG-16 (16-layers CNN developed by the visual geometry group), SVM (support vector machine), QSM (quantitative susceptibility mapping). 
	\end{table}
	
	\noindent 
	\section{Quality Assurance (QA)}
	Brachytherapy QA includes multiple aspects, such as safety, equipment, and plan QA. While the safety and equipment QAs are typically performed manually before the treatment delivery, the plan QA can be performed with the assistance of AI to identify suboptimal plans and improve plan qualities. Three studies used AI models to assist the plan QA, as summarized in Table 9. Reijtenbagh et al. \cite{RN115} trained an RF model using Overlap Volume Histograms (OVHs) to predict the DVH for each OAR in HDR GYN brachytherapy. It predicted doses for the target, bladder, rectum, sigmoid, and small bowels, allowing perform patient anatomy-based QA. This QA model detected sub-optimal plans by identifying when $D_{2cc}$ values fell outsize the 95\% confidence interval, showing strong correlation with MSE ranged between 0.13 and 0.40 Gy. Testing on a different institution’s data, the model identified several clinically compliant plans, proving the model’s effectiveness in multi-center settings. The OVH-based QA model also has advantage in its short training time of less than a minute and short prediction times within seconds. Reijtenbagh et al. \cite{RN78} found an average reduction of 0.62 Gy for all OARs utilizing their QA model, further proving its effectiveness. Another group from EMC \cite{RN79} utilized a fully automated planning model (Erasmus-iCycle) and successfully reduced a mean $D_{2cc}$ of 0.87 Gy in bladder and 1.4 Gy in rectum, yet the study was constrained to single-institute setting. Further, Fan et al. \cite{RN80} developed a DNN model for HDR brachytherapy plan QA, focusing on checking dwell times and positions. The model, based on Inception network by Szegedy et al. \cite{RN81}, used a small number of parameters to reduce the computation burden and the risks of overfitting in traditional CNN methods. Given the non-linear nature of the regression problem, Fan et al. employed a heatmap to represent the probability distribution of the dwell position, reducing the complexity of the task. Predictions of dwell times were within a 2\% deviation from the ground truth, and dwell positions were within one pixel of the planned positions. Despite the long training time of the model, it offered fast and accurate plan QA, predicting dwell times and positions within seconds.
	
	\begin{table}[]
		\caption{AI in QA}
		\label{tab:my-table}
		\resizebox{\textwidth}{!}{%
			\begin{tabular}{|p{2cm}|p{1.5cm}|p{2cm}|p{2cm}|p{7cm}|p{2cm}|}
				\hline
				QA Tasks & Cancer Site & Number of Patients & Model & Result Summary & Citation \\ \hline
				Predict dose-volume histograms & GYN & 145 & RF & MSE between planned and predicted OARs $D_{2cc}$ ranged between 0.13 and 0.40 Gy & [139] \\ \hline
				Treatment Plan QA & GYN & 34 & Erasmus-iCycle & 62 out of 63 plans were comparable or better than clinically generated plans, desired dose obtained in 14 out of 16 plans. & [141] \\ \hline
				Predict dwell positions and times & GYN & 130 & DNN & Dwell times prediction error within 2\% from the ground truth, dwell positions within one pixel of the planned positions. & [142] \\ \hline
			\end{tabular}%
		}
	\end{table}
	
	\noindent 
	\section{Outcome Prediction}
	AI demonstrates proficiency in predicting clinical outcomes, aiding oncologists and physicists can make informed decisions regarding patient treatment. As summarized in Table 10, three different outcome prediction tasks are carried out by machine learning models for both prostate and GYN brachytherapy. Two studies utilized ML methods to make the predictions for locally advanced cervical cancer (LACC) patients. Abdalvand et al. \cite{RN82} studied the effectiveness of four different ML algorithms, LASSO (Least Absolute Shrinkage and Selection Operator) regression, Ridge regression, support vector machine (SVM), and RF in terms of LACC outcome prediction. The ten most important features when considering LACC outcomes were selected from LASSO algorithm to reduce complexity and mitigate over-fitting. The study specifically incorporated patient-specific applicator insertion geometries from 3D MR or CT imaging, as well as the physical, dosimetric, radiobiological, and clinical factors when training the models. The RF algorithm had the highest discriminating ability with an area AUC of 0.82, but an accuracy of only 0.77, limited by the small patient sample (111 selected) and the traditional ML model's ability to address heterogeneities of the clinical data. Tian et al. \cite{RN83} devised a nonlinear kernel based SVM classifier to predict fistula formation from the high radiation doses in patients undergoing interstitial brachytherapy for LACC. Using sequential backward selection and sequential floating backward selection methods, Tian et al. identified 7 most crucial features for model training. Consequently, their model achieved an AUC of 0.904, with sensitivity and specificity rates of 97.1\% and 88.5\%, respectively. 
	
	The recurrence rate of prostate cancer after the initial treatment can be predicted with ML models, offering potential guidance in identifying patients who would benefit the most from salvage HDR brachytherapy. Valdes et al. \cite{RN85} used MediBoost (a decision-tree based model built by the same group \cite{RN84}) and a classification and regression tree (CART) model to predict the 5 years recurrence rate. Only the most important features were selected such as the percentage of positive cores after biopsy and disease-free interval after the first definitive treatment to reduce complexity and risk of over-fitting of the data. The study found a 5-year recurrence probability of 0.75 associated with positive cores >= 0.35 and a disease-free interval < 4.12 years. Although the study was limited by the small dataset of 52 patients, there is 70\% probability that the results were not due to random variation.
	
	\begin{table}[]
		\caption{AI in brachytherapy outcome prediction}
		\label{tab:my-table}
		\resizebox{\textwidth}{!}{%
			\begin{tabular}{|p{2cm}|p{1.5cm}|p{2cm}|p{2cm}|p{7cm}|p{2cm}|}
				\hline
				Application & Cancer Site & Number of Patients & Model & Result Summary & Citation \\ \hline
				Cervical cancer outcome prediction & GYN & 111 & LASSO, Ridge, SVM, and RF & Best AUC = 0.82, sensitivity = 0.79, specificity = 0.76, and accuracy = 0.77 & \cite{RN82} \\ \hline
				Predict fistula development from high radiation dose & GYN & 35 & Nonlinear kernel-based SVM classifier & AUC = 0.904, sensitivity = 97.1\%, and specificity = 88.5\% & \cite{RN83} \\ \hline
				Select patients to receive salvage HDR brachytherapy after first recurrence after radiation therapy & Prostate & 52 & CART and MediBoost & Positive cores >= 0.35 and a disease-free interval < 4.12 years has a second recurrence rate of 0.75, the conclusion has a 70\% probability of not due to random variations & \cite{RN85} \\ \hline
			\end{tabular}%
		}
	\end{table}

	\noindent 
	\section{Real-time Monitoring }
	While various devices and systems can achieve real-time monitoring of HDR brachytherapy, AI offers an alternative solution that enhances quality and efficiency, as detailed in Table 11.
	
	Malignant tumors frequently exhibit distinct temperature distributions compared to normal tissue. Therefore, thermal imaging can aid in identifying tissue-specific changes in the cervix during brachytherapy that are necessary for plan adjustments. Hoffer et al. \cite{RN86} used the k-means method to predict the status of the cervix before and after brachytherapy, using skewness and entropy levels from the thermal image of the cervix. Additionally, K Nearest Neighbors (KNN) and SVM validated a 100\% detection rate for structural and textual changes in cervical tumors before and after brachytherapy. 
	
	During HDR brachytherapy treatments, monitoring the source position and dwell times of the radioactive source can be achieved through a gamma camera, which often suffers from blurring effect and noise. Nakanishi et al. \cite{RN87} proposed a DL based approach to estimate the actual image without blurring effect and noise for better real-time monitoring of the Ir-192 source. The method consisted of two P2P models, similar to the method provided by Isola et al. \cite{RN55} The highest structural similarity index measure (SSIM) value was $0.98 ± 0.006$ and the lowest MAE was $2.2×10^{-3} ±  1×10^{-3}$ Additionally, the full width at half maximum (FWHM) of the estimated image in both horizontal and vertical directions differed by less than 0.5 mm from the actual source size. 
	
	Finally, AI can be used in the calibration of plastic scintillation detectors in a multi-point configuration (mPSD), which provides in-vivo dosimetry measurement and real-time source tracking. Rosales et al. \cite{RN88} trained and compared three algorithms, linear regression, RF, and ANN to calibrate mPSD for real-time feedback in HDR brachytherapy. Among the tree algorithms, the RF had the most accurate calibration result with dose deviation generally remained below 20\% and below 2\% when same range of distances was used for calibration. This ML-based method requires only one calibration for the detector but can still be time-consuming if calibration at multiple locations is needed. While improvements are still needed, this algorithm can lead to more precise measurements of mPSD and help medical physicist and oncologists with adjusting treatment plans.
	
	\begin{table}[]
		\caption{AI in brachytherapy real-time monitoring}
		\label{tab:my-table}
		\resizebox{\textwidth}{!}{%
			\begin{tabular}{|p{2cm}|p{1.5cm}|p{2cm}|p{2cm}|p{7cm}|p{2cm}|}
				\hline
				Monitoring Application & Cancer Site & Number of Patients & Model & Results & Citation \\ \hline
				Monitor thermal image to predict structural and textual changes in cervical tumors before and after brachytherapy & GYN & 6 & k-means, KNN, and SVM & 100\% detection rate for physiological changes in cervical tumors before and after brachytherapy & \cite{RN86} \\ \hline
				Reduce blurring and statistical noise in real-time monitoring of Ir-192 sources with gamma camera & GYN & 11 & Pix2pix models & The highest SSIM = 0.98 ± 0.006 and MAE = $2.2x10^{-3}\pm 1.0x10^{-3}$. FWHM error  0.5 mm in both horizontal and vertical directions. & \cite{RN87} \\ \hline
				Calibration of mPSD & GYN and Prostate & 1936 dwell positions & Linear regression, RF, and ANN & Dose predictions accuracy within 7\% of the TG-43 U1 formalism with all models and less than 2\% deviations using RF model & \cite{RN88} \\ \hline
			\end{tabular}%
		}
	\end{table}
	
	\noindent 
	\section{Discussion}
	In this review, we summarized AI’s applications in different parts of brachytherapy workflows, highlighting its comparable performances to manual efforts in the segmentation, classification, and prediction tasks with non-significant errors. 
	
	The imaging section has the most studies, this is because AI’s role in brachytherapy largely relies on imaging tasks. With the advent of deep learning, image registration has not only become more efficient but also more feasible, especially in terms of achieving precise multi-modal registration. Fast and accurate multi-model image registration is particularly valuable in brachytherapy, where integrating MRI information with TRUS can provide better guidance during the procedure. Besides MRI-TRUS fusion, AI can also enhance resolution and reduce artifacts, providing an alternative solution to overcome imaging challenges in brachytherapy.
	
	Manual contouring is time-consuming and requires extensive training, while AI models can learn segmentation techniques in hours or less. Most studies reported AI segmentation is dramatically faster than human experts on all image modalities. Kraus et al. \cite{RN138} demonstrated that using AI in clinical workflows reduces and standardizes the time from contouring to approval, with AI workflows taking 71–111 minutes, compared to the large deviation of 29–304 minutes with manual workflows. AI's efficiency stems not only from its rapid processing capabilities, but also from the absence of clinical interruptions that human physicians frequently encounter. The reviewed studies showed that AI-generated contours often closely resemble those produced manually in quantitative evaluations. Several other studies have also shown AI’s superior segmentation from different perspectives. For instance, Sanders et al. \cite{RN15} evaluated the differences in dosimetric parameter between the prostate and OAR contours generated solely by a deep-learning algorithm (MIM-Symphony) and those further refined by physicians, finding no significant differences in the dosimetric parameters even after human refinement. King et al. \cite{RN89} conducted a survey on prostate contours on TRUS images generated by AI versus by professionals. The AI algorithm, using nnU-Net, generated contours with median DSC of 0.92 and won a median of 57.5\% of clinical observer preference, showing strong performance both objectively and subjectively. In addition, AI can standardize contours and reduce inter-observer variability errors. As shown in Andersén's \cite{RN25} study, their deep learning model produced results with errors smaller than inter-observer variability. 
	
	Studies on AI applications in treatment planning suggest that AI can generate plans that are comparable to, or even superior to, those created by human planners. This capability stems from AI's ability to learn from hundreds of high-quality treatment plans during its training phase. While manual planning can indeed produce excellent results, it is heavily reliant on the planner's experience and is often resource-intensive and time-consuming. In contrast, AI can generate superior plans much more efficiently. The plan check function from automatic dose prediction models can also detect sub-optimal plans that require proper adjustments.
	
	AI application in applicator reconstruction, particularly multi-needle digitization, offer the potential to facilitate real-time treatment planning and dosimetric adjustments, mitigating discrepancies between the planning stage and the operating room, although there are still challenges in ground truth contour accuracy and needle trajectory prediction. AI models have the potential to simplify more advanced procedures, such as MRI-guided brachytherapy. Overall, AI can improve brachytherapy workflow efficiency, producing similar results to conventional methods in significantly less time with more consistency. 
	
	The outcome prediction, QA, and real-time monitoring sections have the least AI application for several reasons. While AI can assist with brachytherapy preparation, the treatment quality heavily depends on the skills of the physicians or surgeons in the operation room, and physicists responsible for monitoring source delivery and performing QA. The human-centered nature of QA and real-time monitoring limits AI integration in these areas. Outcome prediction, however, is particularly challenging due to the need for long-term data, which is often limited. This lack of comprehensive patient data makes it difficult to apply AI effectively in predicting outcomes, as the reviewed studies only included 35, 52, and 111 patients. 
	
	Two common limitations for the current application of AI in brachytherapy are the small patient cohorts and single institution setups. Analyzing patient numbers involved in each study, we found the mean and the median numbers to be 149 and 66, respectively. A histogram in Figure 9 illustrates the distribution, showing skewness due to a few studies with exceptionally large patient cohorts, while most studies have less than 200 patients. Despite the advantages of brachytherapy, such as shorter overall treatment time and positive outcomes compared to surgery or EBRT \cite{RN112}, there has been a general decline in its use \cite{RN114}.  This is due to procedural difficulty and a reduction in residency brachytherapy training, which has decreased the number of physicians competent in performing this procedure \cite{RN111}.
	
	\begin{figure}
		\centering
		\noindent \includegraphics*[width=6.50in, height=4.20in, keepaspectratio=true]{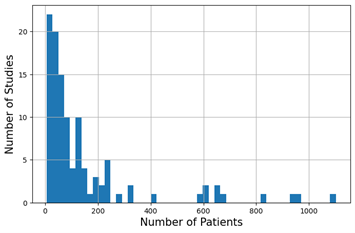}
		
		\noindent Figure 9. Distribution of Number of Patients in investigated studies.
	\end{figure}

	Many studies addressed this challenge by employing cross-validation methods, in which the total available data is partitioned into several groups, with one group for validation and the rest for training. The training of the model ends when each group is used as the validation set once. The cross-validation methods allow the maximum utilization of available data for training and validation by assessing the model's performance across different subsets and enhancing its reliability and generalizability. Data augmentation, which involves applying elastic deformations or transformation such as rotating and resizing, allows different U-Net-based models to gain more training dataset and learn invariance to deformations. Besides, transfer-learning can also be used to leverage the challenge of limited data size, several studies \cite{RN64, RN134, RN14, RN3, RN72} showed that fine-tuning with limited clinical data could lead to more precise task outcomes although requiring longer training times. 
	
	Although various methods exist to mitigate the adverse effects of limited image data, addressing the challenge posed by the lack of patient anatomy and setup variations remains difficult. For example, Holupka et al. \cite{RN102} used 950 training US images from 10 patients, resulting in large seed location errors as large as 2.29 mm when evaluated on new patient sets due to lack of patient variation. AI models may struggle anatomies deviating significantly from training datasets, making reliance on AI-generated results risky, particularly in HDR settings. Diverse data, in terms of various tumor sites, heterogeneities, shielded applicators, and different radiation sources, is crucial for obtaining a model suitable for general-propose brachytherapy dosimetry \cite{RN75}. 
	
	Another significant issue in current AI studies is the reliance on training data from single institutions. Brachytherapy procedures are heavily dependent on imaging protocols, and AI models trained on data from one institution may not perform well elsewhere. Although there were a few studies \cite{RN43, 78} showed no statistically significant difference in cross-institution testing, majority of studies lack this validation. Even using the same imaging protocol, applicator types for brachytherapy could vary, further complicating the implementation of pretrained models \cite{RN70, RN78}. For example, in the applicator reconstruction of cervical cancer brachytherapy section, most of the applicator reconstruction studies used T\&O applicators, but the studies on other types of applicators, such as the T\&R and fletcher applicators, could be further investigated. Although the geometries of applicators may be somewhat similar, it is challenging to apply a model trained on one type of applicator to another type. Dose prediction tasks can be complex due to applicator variability, typically involving one type of applicator setup per study. In addition, the acceptability of AI predicted plans is heavily reliant on the quality of clinical ground truth, influence by variability in treatment planning system and protocol adopted in each institution. Therefore, future generalized AI models development should include large and diverse patient datasets from multiple institutions to address these limitations \cite{RN115}. 
	
	One potential solution for these common issues is to implement foundation models, especially in the field of imaging. Segment Anything Model (SAM) \cite{RN99}, introduced by the Meta AI Research team, is a foundation model for segmenting any object in the input image by detecting valid masks within images and operating through a supervised routine rather than relying solely on unsupervised learning. SAM was pre-trained on a comprehensive dataset with over 1 billion masks sourced from 11 million images. This extensive training employs a task that promotes robust generalization, enabling zero-shot adaptability to unfamiliar objects and images without necessitating supplementary training. Following the introduction of SAM, Ma et al. introduced MedSAM \cite{RN100}, a refined foundation model that specifically integrates medical knowledge to segment medical images. It is designed for universal segmentation tasks and trained on a vast and diverse dataset of over a million medical image-mask pairs that cover 15 modalities and 30 cancer types. MedSAM addresses the issue of the task-specific nature of previous models, which are only capable of segmenting certain image modalities or specific tissues. The diversity in data sources effectively eliminates the need for additional training, addressing the limitation of large biomedical image datasets. MedSAM’s capability to manage various anatomical structures and imaging modalities makes it a potential tool for brachytherapy imaging applications. Large foundation models such as MedSAM can be integrated with fine-tuning methods for more accurate organ segmentation \cite{RN141}, which is applicable to brachytherapy procedures. Future studies aiming to automate the segmentation of organs, tumors, or applicators can achieve this by fine-tuning these models, even with limited, single-institution datasets, potentially leading to more reliable results that are acceptable clinically. Since many aspects of brachytherapy, such as image registration, treatment planning, and applicator placement, depend heavily on accurate image segmentation, the implementation of large foundation models could offer a possible solution to the current challenges in applying AI to brachytherapy. 
	
	The final issue with current implementation of AI is the lack of universal standards in evaluating the performance of AI models. While image segmentation has some commonly accepted evaluation methods, other aspects lack clear reporting guidelines, making it difficult to compare the results of different AI models. Without consistent metrics for comparison, identifying areas for improvement and proposing clear solutions for future studies becomes challenging. The establishment of universal evaluation guidelines for each procedure category is necessary to ensure the future improvement of AI application, thereby enhancing brachytherapy treatments.
	
	\noindent 
	\section{Conclusion}
	This review covers the current development of AI in brachytherapy, focusing on prostate and GYN cancer treatments. Unlike EBRT, which benefits from extensive patient data and straightforward AI implementation, brachytherapy requires additional manual procedural accuracy, introducing unique difficulties for AI integration. Current challenges include the lack of patient data from diverse institutions, and a possible solution is to adopt new foundational models to enhance image segmentation, which paves the way for improvements in other procedures as well. Establishing universal standards in validating the application of AI in brachytherapy is also essential to improve AI’s performance. Despite these challenges, AI has the potential to enhance image segmentation accuracy, provide high-quality plans, simplify real-time planning, which significantly enhance the brachytherapy workflow and encourage the use of brachytherapy. By addressing these challenges and investing in further development, AI can improve the quality of brachytherapy treatment and patient outcomes in the foreseeable future.
	
	\noindent 
	\bigbreak
	{\bf ACKNOWLEDGEMENT}
	
	This research is supported in part by the National Institutes of Health under Award Number R01CA215718, R01DE033512, R01CA272991 and P30CA008748.

	\noindent 
	\bigbreak
	{\bf Disclosures}
	
	The authors declare no conflicts of interest.

	\noindent 
	
	\bibliographystyle{plainnat}  
	\bibliography{arxiv}      
	
\end{document}